\documentclass[pra,aps,twocolumn,longbibliography, nofootinbib, showpacs]{revtex4-2}
\usepackage[colorlinks=true, citecolor=blue, urlcolor=blue, linkcolor = blue ]{hyperref}
\usepackage{graphicx}
\usepackage{bm}
\usepackage{amsmath,amsfonts}
\usepackage{amsthm}
\usepackage{hyperref}
\usepackage{xcolor}
\usepackage{braket}
\theoremstyle{plain}
\usepackage{color}
\usepackage{amssymb}
\usepackage{amsthm}
\usepackage{amsfonts}
\usepackage{float}
\usepackage{tabularx}
\usepackage{graphicx}
\usepackage{multirow}
\usepackage{array}
\usepackage{ragged2e}
\usepackage{booktabs}

\usepackage{mathtools}
\usepackage{esvect}
\usepackage{wrapfig}
\usepackage{amsthm}
\usepackage{verbatim}
\usepackage{bbm}
\usepackage[normalem]{ulem}

\usepackage{enumitem}
\usepackage{fmtcount}
\usepackage{booktabs}
\usepackage{csquotes}
\usepackage{epsfig}

\usepackage{tabularx}
\usepackage{graphicx}
\usepackage{amsmath}
\usepackage{braket}
\usepackage{latexsym}
\usepackage{bm}
\usepackage{graphics,epstopdf}
\usepackage{enumitem}
\usepackage{fmtcount}
\usepackage{booktabs}
\usepackage{csquotes}
\usepackage{epsfig}
\begin{document}

\title{Quantum sensing with ultracold simulators in lattice and ensemble systems: a review}
\author{Keshav Das Agarwal, Sayan Mondal, Ayan Sahoo, Debraj Rakshit, Aditi Sen(De), Ujjwal Sen}

\affiliation{Harish-Chandra Research Institute, A CI of Homi Bhabha National Institute, Chhatnag Road, Jhunsi, Allahabad 211019, India}

\begin{abstract}
    Sensing of parameters is an important aspect in all disciplines, with applications ranging from fundamental science to medicine. Quantum sensing and metrology is an emerging field that lies at the cross-roads of quantum physics, quantum technology, and the discipline in which the parameter estimation is to be performed. While miniaturization of devices often requires quantum mechanics to be utilized for understanding and planning of a parameter estimation, quantum-enhanced sensing is also possible that uses paradigmatic quantum characteristics like quantum coherence and quantum entanglement to go beyond the so-called standard quantum limit. The current review hopes to bring together the concepts related to quantum sensing as realized in ensemble systems, like spin ensembles, light-matter systems, and Bose-Einstein condensates, and lattice systems, like those which can be modeled by the Bose- and Fermi-Hubbard models, and quantum spin models. 
\end{abstract}

\maketitle
\tableofcontents

\section{Introduction}
Metrology is the art of crafting the perfect question and extracting maximal available information from a system. In quantum metrology, the probe state and the measurement strategy play central roles in determining how effectively a physical parameter can be estimated. In quantum-enhanced parameter estimation, the optimal measurement strategy utilizes quantum resources, such as entanglement, squeezing, quantum coherence, and many-body cooperative phenomena, e.g., a probe state being at or near a quantum phase transition. The right strategy can enable precision beyond classical limits, that is the so-called standard quantum limit (SQL),  approaching or even achieving the quantum Cramér-Rao bound \cite{Helstrom1967, Holevo1973, Braunstein1994}, which is the ultimate ceiling for estimating an unknown parameter. Since the idea of quantum sensors was first conceived \cite{Giovannetti2004}, there have been a plethora of theoretical proposals and experimental works that employ quantum systems to estimate various physical quantities ranging from magnetic fields, AC and DC fields, time, pressure, etc., to the detection of gravitational waves and dark matter. The rapid development has quickly established quantum sensing as one of the most mature topics in the field of quantum technology. Interestingly, a large body of effects is within experimental reach with currently available technology. 

A key purpose in quantum metrology is to investigate, identify, and design the most precise possible quantum sensors on physical platforms that are also resilient to external noise. This immediately links quantum sensing with programmable quantum materials. Atomic, molecular, and optical (AMO) systems are among the frontrunners in the realization of quantum technologies, 
encompassing quantum computers, quantum metrology, and quantum cryptography, among others  \cite{bloch2008many,lewenstein2012ultracold,lewenstein2007ultracold,safronova2018search}. The first experimental realization of Bose-Einstein condensation (BEC) in dilute vapours of alkali atoms
has revolutionized the field of ultracold atoms \cite{anderson1995observation,davis1995bose,bradley1995evidence} in the true sense and has quickly established  the AMO systems as a leading subject of research in modern science. 

AMO systems can be prepared and manipulated with unprecedented control via novel techniques. 
Starting from the deterministic preparation of a few atoms in a single micro-trap \cite{serwane2011deterministic,zurn2012fermionization}, bulk systems can be trapped in a magnetic or an optical trap. The role of different statistics naturally comes into play on the AMO platform, as it can feature bosonic and fermionic particles, and 
multicomponent gas mixtures (Bose-Bose, Bose-Fermi, Fermi-Fermi) as well. The internal structure of atoms can be exploited for realizing spin qubits and qudits. Cold atoms and molecules can be trapped in optical lattices created by interfering laser beams, and various spatial configurations in 1D, 2D, and 3D can be generated.  Exciting experimental progress includes single-site and single-atom-resolved advanced imaging techniques \cite{bakr2009quantum,sherson2010single,haller2015single,cheuk2015quantum} and  deterministic preparation of large-scale arrays by individually controlling cold atoms \cite{barredo2016atom}. The intersite tunnelling rates and interatomic interaction can be precisely controlled by tuning optical lattice depth and via Feshbach resonance \cite{chin2010feshbach}. {\color{black}Ultracold atoms are trapped in ultrahigh vacuum,
which reduces their chance of collisions with background gas. This makes them nearly isolated from external environment. In case of optical lattices, the laser interference
creates essentially clean, defect-free lattices. As a result, such systems have a long coherence time, which is orders
of magnitude larger than that of solid systems, because they are largely free from unwanted noises. This makes
them ideal for quantum simulation and a powerful platform for investing in exotic condensed matter phenomena, starting from the observation of Bloch oscillations \cite{ben1996bloch} to the direct measurement of the Zak phase in topological Bloch band \cite{atala2013direct}. Seminal experimental works with cold atoms have successfully demonstrated topological phase transition in the Haldane \cite{jotzu2014experimental}, Harper-Hofstadter \cite{aidelsburger2013realization}, and SSH models \cite{meier2016observation} by engineering synthetic dimensions \cite{ArgelloLuengo2024}, spin-orbit coupling \cite{lin2011spin,wang2012spin} and Floquet dynamics \cite{eckardt2017colloquium}. }

The early proposals for quantum-enhanced sensing via Ramsay interferometry demand preparation of multiparty-entangled GHZ-type states to encode the unknown parameter through a unitary phase shift. Such entangled probe states result in Heisenberg scaling that allows precision to scale with the number of atoms. GHZ-type states have been realized across AMO platforms, utilizing techniques like Rydberg blockade in cold atom arrays \cite{pachniak2021creation}, spin exchange dynamics in bosonic ensembles \cite{monz201114}, and high-fidelity entangling gates in trapped-ion systems \cite{Chen2024}. These protocols are, however, less robust against perturbations. Despite enormous successes, preparation of decoherence-free large atomic ensembles poses technical challenges and is awaiting next-generation ideas. Another route for improved and relatively robust parameter estimation is via a collective spin ensemble that naturally embodies the symmetry and spin structure of the well-known Lipkin-Meshkov-Glick (LMG)-type models, featuring all-to-all interactions amongst spins. Cavity QED set-ups and collective spin models in Bose-Einstein condensates, where effective long-range interaction is created by photon-mediated interactions, can simulate the LMG models and quantum phase transitions in such systems 
\cite{morrison2007collective,huang2011quantum,Muniz2020}.

Recently, paradigmatic condensed matter models have opened new avenues in quantum sensing, facilitating the design of sensors with enhanced precision and resilience by exploiting quantum phenomena, such as spin squeezing \cite{hald1999spin,Ma2011} and quantum criticality, characterized by gap closing, symmetry-breaking, and long-range correlations \cite{reviewMontenegro2024}. These typical short-ranged quantum many-body model Hamiltonians, which include quantum spin chains (XY, Heisenberg models), Bose- and Fermi-Hubbard models can be routinely realized  in the AMO platforms via ultracold atoms and molecules in optical lattice, trapped-ions, and Rydberg atoms in tweezer arrays \cite{Greiner2002,esslinger2010fermi,porras2004effective,korenblit2012quantum,Monroe2021}. 
Finite-range interactions with metrological relevance can also be engineered by employing dipolar atoms and molecules,  laser-excited Rydberg atoms, or trapped ion quantum simulators \cite{Su2023, Bao2023}. The idea of quantum critical sensors, which falls under the broad category of adiabatic quantum many-body sensors, is not limited to the second-order quantum phase transitions but spans those beyond the Landau paradigm, e.g., topological and localization transitions. These bring exciting new prospects, given the fact, as mentioned previously, that seminal cold atom experiments have successfully demonstrated topological phase transitions in the Haldane and the SSH models. Similarly, landmark cold-atom experiments have successfully explored single-particle localization and many-body localization in the presence of interaction by using highly controllable potentials in the form of speckled disorder \cite{Billy2008} or quasiperiodic potentials through bichromatic optical lattices \cite{Schreiber2015}. In addition to these, novel methods beyond conventional ones are also being developed for quantum sensing~\cite{Zhang2022, PhysRevA.106.062442, PhysRevLett.132.240803,PhysRevLett.134.110802}.  

Moreover, long coherence times in AMO and ion-trap platforms have allowed precise studies of many-body dynamics, such as quenches, Floquet phases, and time crystals, using techniques such as lattice shaking, Rydberg interactions, and tunable spin-spin couplings \cite{Huang2018,Chinzei2020, Muniz2020}. 
Such techniques are paving the way for a new kind of quantum sensors--the many-body dynamical quantum sensors, where time is an additional resource along with system size. 

There are several extremely useful reviews on quantum sensing and metrology, covering specific topics from foundations to practical applications. Refs.~\cite{Paris2009, Mukhopadhyay2025, Pezze2025} reviews foundational aspects and theoretical development of quantum sensing and metrology. Refs.~\cite{Degen2017, Huang2024} discusses quantum metrology in various quantum systems with Ref. \cite{reviewMontenegro2024} focused on quantum metrology with quantum many-body systems. Refs.~\cite{Huang2014, Pezze2018, Szigeti2021} provides  earlier developments of metrological aspects in cold atoms. Refs.~\cite{Robins2013, Geiger2020} and Ref.~\cite{Mehboudi2019b} cover inertial sensors and thermometry, respectively. The current review offers complementary viewpoints to these critical resources.  In particular, in this review, we attempt to assemble quantum metrological protocols relevant to the AMO platforms and aim to present brief details on the important works in this arena. In the following, Sec.~\ref{sec-qfi} presents a necessary theoretical framework of quantum metrology. Sections~\ref{sec-ensemble} and~\ref{sec-lattice} discuss the ensemble and lattice systems, respectively. This is followed by a brief discussion in the Sec.~\ref{sec-disc}.

\section{Quantum sensing and its connection to Fisher information}
\label{sec-qfi}
The central task in quantum sensing is to estimate a single or multiple parameters in a system with the highest attainable precision, obtained through various measurements~\cite{Helstrom1967, Giovannetti2004, Paris2009}. 
In particular, it is used to infer the unknown parameter(s) with maximum accuracy from the observed data, minimizing the variances. Precisely, suppose the parameters $\theta$~\footnote{$\theta$ is denoted for multiple parameters, $\theta_i$.} are encoded in a physical system, which gives a probability distribution $p_\theta(x)$ of measurement outcomes $x$~\footnote{$x=\{x_i\}$ represents the set of outcomes obtained for the set of parameters, $\{\theta_i\}$.}. The task is to find the function of the outcomes, i.e. an estimator $\hat{\theta}(x)$ that gives the value of $\theta$ with best precision. For example, the mean of a Gaussian distribution, $p_\theta(x)\sim e^{\frac{-(x-\theta)^2}{\sigma}}$ can be guessed by $N$ identical independent measurements, with $\hat{\theta}(x)=\frac{1}{N}\sum_i x_i\equiv\bar{x}$ as an estimator, with precision $\delta\hat{\theta}(x) = \sqrt{\frac{\sum_i(x_i - \bar{x})^2}{N}}$. The estimation theory generalizes this analysis for arbitrary measurements and an estimator, using a key quantity, called Fisher information~\cite{Holevo1973, Braunstein1994} $F(\theta)$, of the given encoding $p_\theta(x)$.

\subsection{Fisher information}
Consider a statistical model where data $X$ is distributed according to a probability density $p_\theta(x)$, parameterized by $\theta$, which is to be estimated. In the statistical manifold of the probability distributions, $\theta$ acts as a coordinate whose infinitesimal displacements $d\theta$ gives the notion of distance between neighboring distributions. This can be computed by Kullback-Leibler divergence $D_{\mathrm{KL}}$ (or the relative entropy) between two distributions $p_\theta(x)$ and $p_{\theta+d\theta}(x)$, with $d\theta$ being a perturbation, given by
\begin{align}
    D_{\mathrm{KL}}(p_\theta(x) \| p_{\theta+d\theta}(x)) &=\int p_\theta(x) \ln \frac{p_\theta(x)}{p_{\theta+d\theta}(x)} dx \nonumber \\
    &\approx \frac{1}{2}F(\theta)d\theta^2,
\end{align}
where $F(\theta)=\mathbb{E}\left[\left(\mathrm{d}_\theta[ \ln p_\theta(x)]\right)^2\right]$ is the Fisher information, with $\mathrm{d}_\theta\equiv\frac{\mathrm{d}}{\mathrm{d}\theta}$ and the expectation $\mathbb{E}[\cdot]$ is taken over the $p_\theta(x)$ probability distribution. Therefore, given a random variable $X$, dependent on an unknown parameter $\theta$, the Fisher information quantifies the information content about $\theta$, carried out by
by its underlying probability distribution.

As the optimal estimator can be parameter-dependent, the optimal metrological protocol increases the precision of estimation over an overall guess of the parameter value $\theta_0$. Specifically, a function $\hat{\theta}(x)$ of measurement outcomes $x$, is locally unbiased around $\theta\sim\theta_0$, when 
\begin{equation}
    \begin{aligned}
\mathbb{E}\left[\hat{\theta}(x)\right]_{\theta=\theta_{0}} & =\int \mathrm{d} x \:\hat{\theta}(x) p_{\theta_{0}}(x)=\theta_{0}, \quad \text{and} \\
\left.\mathrm{d}_\theta\left(\mathbb{E}\left[\hat{\theta}(x)\right]\right)\right|_{\theta=\theta_{0}} & =\left.\int \mathrm{d}x\: \hat{\theta}(x) \mathrm{d}_\theta [p_{\theta_{0}}(x)]\right|_{\theta=\theta_{0}} = 1,
\end{aligned}
\end{equation}
{\color{black} with $\mathrm{d}_\theta\equiv\frac{\mathrm{d}}{\mathrm{d}\theta}$ is the derivative with respect to the parameter $\theta$ to be estimated.}
The Cramér-Rao gives a fundamental bound on the variance of any locally unbiased estimator $\hat{\theta}(x)$\footnote{$\delta^2\hat{\theta}:=\mathbb{E}\left[\hat{\theta}^2(x)\right]-\left(\mathbb{E}\left[\hat{\theta}(x)\right]\right)^2$} as $\delta^2\hat{\theta}\geq \frac{1}{F(\theta)}$. It generalizes to multi-parameter estimation, i.e. estimating $\{\theta_1, \theta_2,\dots\theta_m\}\equiv\vec{\theta}$, via Fisher information matrix as $F^{ij}(\vec{\theta}) = \mathbb{E}\left[\partial_{\theta_i} [\ln p_{\vec{\theta}}(x)]\partial_{\theta_j} [\ln p_{\vec{\theta}}(x)]\right]$, and covariance matrix $C^{ij}=\mathbb{E}\left[\hat{\theta}_i\hat{\theta}_j\right]-\mathbb{E}\left[\hat{\theta}_i\right]\mathbb{E}\left[\hat{\theta}_j\right]$. The Cramér-Rao bound gives $C-F^{-1}\geq 0$, i.e. is a positive semi-definite operator. Here, the parameter(s) $\theta$ is considered as a constant, known as the frequentist approach, and Cramér-Rao bound can be achieved asymptotically with increasing number of measurements. With few measurement outcomes, Bayesian estimation, where $\theta$ can be regarded as a random variable, provides a tighter Bayesian Cramér-Rao bound, which matches with the frequentist approach with increasing number of measurements. In this review, we will focus on the frequentist approach in quantum metrology and sensing in ultracold systems.

Increasing Fisher information $F(\theta)$ provides better attainable bound of precisions. Using measurements on $N$ systems, $F(\theta)$ increases with increasing $N$, and its scaling denotes the metrologically useful properties of the underlying systems. As Fisher information is an additive quantity, $F(\theta)\propto N$ is the best scaling for classical systems, with variance $\delta^2\hat{\theta}\propto N^{-1}$ and standard deviation $\delta\hat{\theta}\propto N^{-\frac{1}{2}}$, which is known as the shot-noise limit (SNL).

\subsection{Quantum Fisher information}
The states of quantum systems are represented by density operators $\rho$, on which the parameter $\theta$ that is to be estimated, is encoded and a measurement is a set of positive semi-definite operators $\{E_k\}$, with $\sum_k E_k=\mathbb{I}$ (identity), which is termed as positive operator valued measurement (POVM), is performed. 
To estimate the parameter $\theta$, suppose a measurement is performed on the encoded state $\rho(\theta)$ and the outcome obtained is, say, $k$, with the conditional probability $p_{k}(\theta) = \text{Tr}[\rho(\theta)E_k]$. Optimizing over all POVM-based strategy leads to the lower bound on the mean square error as $\delta^2\hat{\theta}\geq \frac{1}{\mathcal{F}_\theta}$, where $\mathcal{F}_\theta$ is the quantum Fisher information (QFI), providing the quantum Cramér-Rao bound.
The optimal measurement choice can be constructed from the eigenbasis of a Hermitian operator $L_\theta$, called the symmetric logarithmic derivative (SLD) operator, satisfying 
\begin{equation}
    \mathrm{d}_\theta \rho(\theta)\!=\!\frac{L_\theta\rho \!+\! \rho L_\theta}{2};\:\text{with} \: \bra{\Psi_k}\!L_\theta\!\ket{\Psi_l}\!=\!\frac{2\bra{\Psi_k}\mathrm{d}_\theta\rho(\theta)\ket{\Psi_l}}{\lambda_{k}+\lambda_{l}},
\end{equation}
in the eigen-decomposition of the encoded state $\rho(\theta) = \sum_k \lambda_k\ket{\Psi_k}\bra{\Psi_k}$ and $\lambda_k+\lambda_l\neq0$. The optimal measured observable $A$ is such that $i(\rho A-A\rho)=\mathrm{d}_\theta[\rho(\theta)]$, which can be a highly non-local in nature. The QFI can be written as
\begin{equation}
    \mathcal{F}_\theta=\text{Tr}[\rho(\theta)L^2_\theta] = 2\sum_{k,l}\frac{|\bra{\Psi_k}\mathrm{d}_\theta\rho(\theta)\ket{\Psi_l}|^2}{\lambda_{k}+\lambda_{l}}.
    \label{eq:qfi_def}
\end{equation}

Beyond single parameter estimation, when multiple parameters, $\vec{\theta}=\{\theta_1, \theta_2,\dots,\theta_m\}$, have to be estimated, one considers quantum Fisher information matrix, whose elements are 
\begin{equation}
\mathcal{F}_{\vec{\theta}}^{ij}= 2\sum_{k,l=0}^{d-1} \frac{\text{Re}(\bra{\Psi_k}\partial_{\theta_i}\rho\ket{\Psi_l}\bra{\Psi_l}\partial_{\theta_j}\rho\ket{\Psi_k}}{\lambda_{k}+\lambda_{l}},
\end{equation}
where $\rho=\sum\limits_{k=0}^{d-1}\lambda_k\ket{\Psi_k}\bra{\Psi_k}$ is considered to be full rank. 
In this case, the Cramér-Rao theorem states that the inverse of the Fisher information matrix is the lower bound of the covariance matrix $\text{Cov}_{\vec{\theta}}^{ij}=\langle\theta_i\theta_j\rangle-\langle\theta_i\rangle\langle\theta_j\rangle$ \cite{Liu2019}.


If the parameter is encoded in a pure quantum state $\ket{\psi_\theta}$ the QFI in Eq. (\ref{eq:qfi_def}) reduces to 
\begin{equation}
    \mathcal{F}_{\theta}\left[\ket{\psi_\theta}\right] = 4\left[\bra{\mathrm{d}_\theta \psi_\theta}\mathrm{d}_\theta \psi_\theta\rangle-\left|\bra{\psi_\theta}\mathrm{d}_\theta \psi_\theta\rangle\right|^{2}\right].
\end{equation}
When $N$ copies of $\rho(\theta)$ are available, i.e., by using $\rho(\theta)^{\otimes N}$, we have $\mathcal{F}[\rho(\theta)^{\otimes N}]=N\mathcal{F}[\rho(\theta)]$, exploiting the additivity of QFI. It was also shown that even when classical correlations are allowed among $N$ copies of $\rho(\theta)$, $\mathcal{F}\sim N$ \cite{Pezze2009}, known as the standard quantum limit (SQL). On the other hand, entanglement among $\rho(\theta)^{\otimes N}$ can beat SQL scaling, with $\mathcal{F}(\theta)\sim N^2$ being referred to as Heisenberg limit (HL), thereby ensuring quantum advantage in sensing. Further, there are sensing protocols in which $\mathcal{F}_\theta\sim N^\beta$ with $\beta>2$, exhibiting super-Heisenberg scaling.
{\color{black}
\subsection{Difference between SNL and SQL} The SNL and SQL are both usually referred to as having a $1/\sqrt N$ scaling and are often used synonymously. Here, we discuss the difference between the two in certain contexts. In an interferometer, a beam of classical light impinges on the beam splitter and get divided into two parts. These two parts travel along different optical paths, picking up a phase difference which is inferred by measuring the intensity of the output beams. For a coherent light beam with an average photon number $N$, the photons are distributed between the two output ports of the interferometer, depending on the phase difference acquired by the two paths. A fraction of the photons are detected at one of the output ports, while the remaining photons are detected at the other, with the relative proportions determined by the phase difference between the interfering beams. In one of the outputs, the average number of photons detected is $N\cos^2(\theta/2)$ ($N\sin^2(\theta/2)$ is detected in the other output), where $\theta$ is the phase difference picked up by the two beams.  The error in sensing $\theta$, $\Delta \theta$ is given by the inverse of the signal-to-noise ratio ($SNR$) of such a measurement, $SNR \propto \frac{\bar{n}}{\Delta n} = \frac{|\alpha|^2}{|\alpha|} = |\alpha| = \sqrt{N}$. This scaling of $\Delta \theta \sim 1/\sqrt{N}$ is called the shot-noise limit (SNL). This arises due to the Poissonian distribution of photons from coherent sources when considering sensing using interferometric setups. This is true for single-photon beams as well, where, due to the central limit theorem, the error in sensing $\theta$ scales as $1/\sqrt{N}$, where $N$ is the number of photons being used. In addition to optical interferometers, this is observed in atomic interferometers as well, where phase is measured by observing the atomic population in the output of the interferometer.       

The standard quantum limit, or SQL, is the metrological scaling obtained when the system under investigation undergoes procedures which do not fully exploit the quantum nature of the system. As discussed earlier, due to the additive nature of Fisher information, for $N$ probe states, total fisher information $\mathcal{F}_N = N\mathcal{F}$ leading to $1/\sqrt{N}$ scaling for SQL. Sometimes in literature, SQL and SNL are used interchangeably due to similar $1/\sqrt{N}$ scaling, but in certain contexts they can differ.    

In continuous-variable sensing, especially in gravitational wave detection, the SQL is determined by the optimization of two noises, namely, shot noise and radiation pressure noise. The shot noise arises due to the fluctuations in the number of photons detected. On the other hand, when photons reflect off mirrors or test masses, they exert a small force on such objects, leading to a small perturbation. The perturbation caused by a single photon is not significant, but when continuous pulses are sent, the perturbations add up and become significant. This back-action leads to the radiation pressure noise. The shot noise is inversely proportional to the optical power, whereas the radiation pressure is proportional to the power.
The total noise involves both the shot noise and radiation pressure noise, and minimizing one may increase the other. 
In this case, the SQL is obtained in the limit at which the sum of the two noises is minimal~\cite{Giovannetti2004, Danilishin_2012, Danilishin_2019}. The SQL is the sweet spot in this trade-off and can only be beaten by using quantum natured light and quantum non-demolition measurement~\cite{PhysRevD.23.1693, RevModPhys.82.1155}.  
From now on, by either SQL or SNL, we imply $1/\sqrt{N}$-scaling, where $N$ being the number of uncorrelated, independent, and identical probes.
}

\subsection{Parameter encoding}

{\color{black} In general, the parameter dependence of a Hamiltonian can be expressed as
$H(\theta) = H_s+\theta H_e$, where $H_s$, is the parameter-independent control Hamiltonian of the system Hamiltonian and $H_e$ is the parameter-encoding term that imprints information of $\theta$ on the quantum probe state $\rho_0$. Therefore, the parameter $\theta$ can be encoded via various different methods based on the choice of $H_s$ and $H_e$. Two generically used protocols are discussed below.}

\subsubsection{Interferometry-based sensing}
\label{interferometry_squeezing}
 Quantum interferometry is the most commonly used technique for quantum sensing~\cite{Giovannetti2004}, in which the parameter to be estimated is encoded in a unitary operation governed by a Hamiltonian $H(\theta)$, i.e., $U_\theta(t) = e^{-iH(\theta)t}$. 
 Initializing the probe and auxiliary states in $\rho_0$, the system evolves for time $t$ as $\rho_\theta(t) = U_\theta(t)\rho_0U_\theta^\dagger(t)$. Here $U_\theta(t)$ involves the Hamiltonian $H(\theta)$ with the assumption that {\color{black} it depends linearly on parameter $\theta$.} At time $t$, the goal is to estimate $\theta$ by performing optimal measurements on probe and auxiliary systems. Let us first consider the situation {\color{black} without any control Hamiltonian and independent encoding, i.e., $H_s = 0$ and non-interacting $H_e = \sum_{i=1}^N H_i$ with local $H_i = \sum_{k =0}^M E_k|k\rangle\langle k|$ representing identical single systems}. Choosing the initial $N$-party entangled probe state as the cat state, $|\psi_{0M}^N\rangle = \frac{1}{\sqrt 2} (|0\rangle^{\otimes N} + |M\rangle^{\otimes N})$ where $|0\rangle$ ($|M\rangle$) is the eigenstate of $H_i$ having minimum (maximum) eigenvalue $E_0$ ($E_M$), the evolution under the Hamiltonian $H_e$ leads to the variation as $\delta^2 H_e = \langle {H_e}^2\rangle - \langle H_e\rangle ^2 = \sum_i \delta^2 H_i = N^2\frac{(E_M - E_0)^2}{4}$, thereby attaining the HL. On the other hand, preparing all the probe states in $|\psi_{0M}\rangle = \frac{1}{\sqrt 2} (|0\rangle + |M\rangle)$, one can only achieve the spread as $N\frac{(E_M - E_0)^2}{4}$, resulting in SQL and the corresponding uncertainty $\delta^2  \hat \theta \sim \frac{1}{t^2(\delta H_e)^2} = \frac{4}{Nt(E_M-E_0)^2} $  \cite{Giovannetti2004, Boxio2007}. When the Hamiltonian involves  $k$-body interactions, 
 $\mathcal{F}_\theta[\rho(\theta,t)]\sim t^2\Vert H_e\Vert^2$ \cite{Boxio2007}, where $\Vert H_e\Vert$ is the width of the eigenspectrum of $H_e$ and in this case $\Vert H_e\Vert\sim N^k$.

{\textbf{\emph{Collective spins.}} In interferometric setups, the two modes undergo evolutions along two different paths, leading to a phase difference between them, which is then estimated by observing their interference fringes~\cite{Cronin2009}. 
Restricting all the degrees of freedom of the system into two modes one can effectively describe the atom as an arbitrary spin-$1/2$ particles a qubit~\cite{Yurke1986}, where the two modes, $|a_1\rangle$ and $|a_2\rangle$ can be represented as spin-up and spin-down respectively.

The description of ensembles of $N$ distinguishable qubits with the help of the collective spin operator $\vec{\hat J}  = \{J_x, J_y, J_z\}$ where $\hat J_\alpha = \sum_{k=1}^N \sigma_\alpha^{(k)}$ and $\hat{J}^2= \hat{J}_x^2+\hat{J}_y^2+\hat{J}_z^2$ is the square of the total spin operator $\vec{\hat J}$, with $\{\sigma_\alpha\}$ ($\alpha = x,y,z$) being the Pauli matrices.
The three orthogonal spin components follow commutation relation $[\hat J_\alpha,\hat J_\beta] = \epsilon_{\alpha\beta\kappa} \hat J_\kappa$, where $\epsilon_{\alpha\beta\kappa}$ is the Levi-Civita symbol. 
Since the operators $\hat J^2$ and $\hat J_z$ commute, their eigenstates form a basis given by $|j,m\rangle$ , such that $\hat{J}^2|j,m\rangle = {j(j+1)}|j,m\rangle$ and $J_z|j,m\rangle = m|j,m\rangle$, with $0\leq j\leq N/2$ and $-j\leq m\leq j$. The sector with $j=\mathcal{J}\coloneq N/2$, has $m=-\mathcal{J},..,\mathcal{J}$ and consists of $(N+1)$-qubit symmetric states, called the Dicke state~\cite{Dicke1954}, which are common eigenstates of $\hat{J}_z$ and $\hat{J}^2 = \frac{N}{2}(\frac{N}{2}+1)$. By using this $N$-party entangled state, HL can be achieved in quantum metrology.

Let us now introduce $N$-qubit coherent spin states represented by $|\alpha,\varphi\rangle \coloneq [\cos(\alpha/2)|0\rangle + e^{i\varphi}\sin(\alpha/2)|1\rangle]^{\otimes N}$ which is a separable state. 
Let us suppose that in a coherent spin state, all spins are pointing in the $z$-direction, i.e., $J_z = \mathcal{J}$ while the perpendicular directions will have isotropic variances ($\delta J_x^2 = \delta J_y^2$) with $\delta J_x^2 + \delta J_y^2 = \langle \hat J^2\rangle - \langle \hat J \rangle ^2= \mathcal{J}$, since $\delta J_z^2 = 0$ and $\langle \hat J\rangle = \langle \hat J_z\rangle = \mathcal{J}$ leading to $\delta J_x^2 = \delta J_y^2=\frac{1}{2}\mathcal{J}$. Thus, for any arbitrary coherent state, the perpendicular spin fluctuations $\delta J_\perp^2 = \mathcal{J}/2 = N/4$. Hence, the sensitivity of the phase $\theta$ to be estimated is given by  $\delta \hat \theta = \delta \hat J_\perp /(\partial \langle \hat J_\perp\rangle/\partial \theta)$, with $\partial \langle \hat J_\perp\rangle/\partial \theta \sim N$, leading to the SQL,  $\delta \hat \theta \sim 1/\sqrt N$. }

{\textbf{\emph{Spin Squeezed states.}} Spin squeezed states have anisotropic spin fluctuations in the direction perpendicular to the mean spin~\cite{Kitagawa1993, Ma2011}. The fluctuations are decreased in one direction, consequently it increases in the other one, due to the Heisenberg uncertainty principle. Let us discuss three notable quantifiers of spin squeezing. First, the squeezing parameter $\xi_N^2 = \delta \hat J^2_{\perp ,\min} /(\mathcal{J}/2)$ is the ratio of the fluctuations in the direction having minimum value to $\mathcal{J}/2$, with the state being spin-squeezed if $\xi_N^2 < 1$. 
Second, $\xi_R^2 =  (\delta \hat{J}_{\perp, \min})^2/(2\mathcal{J}) $, where the $\delta J_{\perp,\min} \delta J_{\perp, \max} \geq \mathcal{J}/2$, with $\delta J_{\perp , \max}$ being the fluctuation of the other direction. Unlike the first measure, these states have large spin mean length, are useful in standard metrological applications and $\xi_R^{-1}$ measures the precision gain~\cite{Wineland1992, Wineland1994, Ma2011}. Third, $\xi_S^2 = 2\delta \hat J^2_{\perp,\min} / \mathcal{J}$~\cite{Kitagawa1993, Ma2011}, with the states being entangled when $\xi^2_S <1$. 
Let a relative phase $\theta$ be encoded by rotating the initial spin squeezed state by $\theta$ around an axis, say $\hat J_z$. The phase resolution of the interferometer is given by  $\delta \hat \theta^{-1} = \left(\delta \hat J_z /(\partial \langle \hat J_z\rangle/\partial \theta)\right)^{-1}$. The maximum sensitivity is reached when $(\partial \langle \hat J_z\rangle/\partial \theta)_{\max} = \mathcal{V} N/2$, where $\mathcal{V}$ is the visibility, such that $\mathcal{V} \coloneq 2\langle \hat J\rangle/N$. Thus, we get $\delta \hat \theta = \xi_S \frac{1}{\sqrt N}$, by putting $\xi_S = 2\delta \hat J_z/\mathcal{V}\sqrt N$~\cite{Gross2012}, as introduced before but now the visibility factor is included. }

\subsubsection{Criticality-based sensing}
\label{critical_sensor}
In recent times, it has been observed that interacting many-body systems having quantum critical point~\cite{Sachdev2011} of $N$ qubits, can be used as a quantum sensor. 
In particular, at zero-temperature, several quantum spin models undergo a second-order quantum phase transition at quantum critical point at which the gap-closing typically occurs.
Consider a system  governed by the Hamiltonian $H = H_s + \theta H_e$, with $H_s$ and $H_e$ being non-commuting, and the characteristics of the ground state (lowest energy eigenstate) changes its behavior drastically with the variation of $\theta$. In this approach, the parameter $\theta$ is encoded in the ground state which acts as the probe state of the quantum metrology. It has been shown that when the system approaches its critical point, $\theta = \theta_c$, the quantum Fisher information of the ground state scales quadratically with the increase of ths system size. Note that there are quantum spin models which possess multiple criticalities and it was shown that with the system size, QFI $\propto N^\beta$ with $\beta>2$. Furthermore, instead of equilibrium, the dynamical state evolved according to the Hamiltonian $H$ also can provide nonlinear scaling in QFI~\cite{Chu2021, Mishra2021}.

\section{Sensing in ensemble systems }
\label{sec-ensemble}
In this section, we discuss ensemble-based sensing that utilizes the collective quantum response of systems, such as Bose–Einstein condensates,  cavity-coupled cold atoms, strongly coupled light–matter, and  spin ensembles. High-precision measurements can be performed using spin-squeezed states, cavity-mediated interactions, quantum phase transitions, and atomic condensates.
\subsection{Spin ensembles}
The ensemble of spins with collective interactions can be represented via the Lipkin-Meshkov-Glick (LMG) model \cite{Lipkin1965, Meshkov1965, Glick1965}. It is described via the collective spin operator, $\hat{J}$, and the corresponding Hamiltonian is given by
\begin{align}
H_{LMG} ={}&
-\frac{\lambda}{N} \left( \frac{1 + \gamma}{2} \right)
(\hat{J}^{2} - \hat{J}_{z}^{2})
- h \hat{J}_{z} \nonumber \\
&-\frac{\lambda}{2N} \left( \frac{1 - \gamma}{2} \right)
\left( \hat{J}_{+}^{2} + \hat{J}_{-}^{2} \right), \nonumber
\end{align}
where $h$ is the magnetic field strength in the $z$-direction, $\lambda$ denotes the interaction strength, $\gamma$ is the anisotropy parameter, and $N$ are the number of spins in the system. The total spin of the system remains conserved. Additionally, it has spin-flip ($\mathbb{Z}_2$) and time-reversal symmetry.
$\gamma=1$ turns out to be a special case with $\text{U}(1)$ symmetry and conserved $\hat{J}_z$. $H_{LMG}$ can be solved analytically using Bethe's ansatz \cite{Pan1999} or the Holstein-Primakoff transformation \cite{Dusuel2004}, revealing quantum phase transitions with spontaneous symmetry-breaking at $h=1\coloneq h^{cr}_{LMG}$ which can help in improved sensing of the magnetic field $h$, or a parameter $\theta$, encoded dynamically on the ground state via a local generator.

The fidelity susceptibility of the ground state of the LMG model, which is related to the quantum Fisher information (QFI) via a simple multiplicative factor, and its scalings have been extensively studied~\cite{Kwok2008, Ma2009a, Castaos2012}. Consequently, a definitive quantum advantage can be deduced at the critical points in quantum sensing of the magnetic field~\cite{Frerot2018}.  
Recently, these investigations have been extended to include dissipative cases~\cite{Pavlov2023}. 
In the interferometric-based sensing, the studies on non-classical properties of the ground state, which is a spin-squeezed state, are performed in \cite{ Ma2009b}. Specifically, $\xi_N$ and $\xi_R$  are analyzed in relation to the metrological properties of the ground state of $H_{LMG}$. With increasing squeezing near the critical point $h^{cr}_{LMG}$ and unitary encoding of the parameter $\theta$ in the perpendicular direction of squeezing, it is possible to saturate the HL, $\delta\hat{\theta}\sim N^{-1}$ in the isotropic case $\gamma=0$. In the anisotropic case, a sub-Heisenberg-limit scaling $\delta\hat{\theta}\sim N^{-5/6}$ for  $h < h^{cr}_{LMG}$, whereas the HL is achieved in the phase, $h>h^{cr}_{LMG}$.

The criticality-based sensing strategy in the LMG model, particularly focusing on the estimation of anisotropy and the use of LMG systems as quantum thermometers, has been investigated  in~\cite{Salvatori2014}, highlighting that criticality in LMG systems can be leveraged to enhance precision in quantum metrology, potentially achieving ultimate bounds with finite sizes and temperature. It has been shown that a local measurement scheme enables thermal states to attain maximum quantum Fisher information (QFI) at the critical point, and  hence, demonstrates the potential of $H_{LMG}$ in quantum thermometry ~\cite{Ostermann2024}. The closing of the energy gap near criticality causes adiabatic slowing down, which has given rise to the alternative idea of dynamical quantum sensing devices. The metrological properties of the dynamical states under the unitary evolution governed by $H_{LMG}$ has also studied. It has been shown that the dynamical states beat the SNL, $\mathcal{F}_\theta\sim N^{4/3}$ ~\cite{Gietka2022a}. Recently, various sudden quench, adiabatic and finite-time-ramp protocols have been studied, taking into account the time required for the encoding or the state preparation time~\cite{Garbe2022}. The authors show that while HL ($\mathcal{F}_\theta\sim T^4$) is attainable via homodyne measurement in a quadrature, it requires non-standard measurement setups. The isotropic LMG model generates one-axis twisting dynamics~\cite{Li2023}, and has been simulated in bosonic Josephson junctions and Bose–Einstein condensates~\cite{Chen2009, Zhou2023}, which will be discussed later with a focus on their twisting properties. While in $H_{LMG}$, spins interact with all-to-all interactions, such advantages are also seen in the systems with weak pairwise  interaction~\cite{Kuriyattil2025}, particularly in the lattice systems with nearest-neighbour interactions. It will be discussed with brief details in Sec.~\ref{sec-lattice}.

\subsection{Light-matter systems}
In quantum optics, interactions between light and matter via coupling between photons and atoms can lead to non-classical properties, such as entanglement between photonic and atomic (or spin) degrees of freedom.~\cite{Lambert2004, Lambert2005}  and quantum criticality~\cite{Hepp1973, Wang1973, Duncan1974, Emary2003, Castaos2009, Hwang2015, Hwang2016, Shen2017, Ying2021}. Specifically, $N$ atoms, with energy levels modelled with spin degrees of freedom $\hat{S}_\alpha^k (\alpha=x,y,z)$ and $\hat{J}_\alpha=\sum\limits_{k=1}^{N}\hat{S}_\alpha^k$ interact with quantized photons in a cavity, modeled as harmonic oscillator with $\hat{b}^\dagger, \hat{b}$ and $\hat{n}=\hat{b}^\dagger \hat{b}$ as creation, annihilation and number operators, respectively, and can exhibit quantum phase transitions. Interactions are then given by $H^{rt}_{int}=\hat{b}\hat{J}_{+} + \hat{b}^\dagger\hat{J}_{-}$ and $H^{crt}_{int}=\hat{b}\hat{J}_{-} + \hat{b}^\dagger\hat{J}_{+}$ indicating rotating terms and counter-rotating terms, respectively, with $\hat{J}_{\pm}=(\hat{J}_{x}\pm i\hat{J}_{y})/2$. Therefore, the light-matter systems can be described by 
\begin{equation}
    H_{LM} = \hbar\Delta\hat{J}_z + \hbar\omega\hat{n} + \frac{\tilde{\lambda}_1}{\sqrt{N}}H^{rt}_{int} + \frac{\tilde{\lambda}_2}{\sqrt{N}}H^{crt}_{int},
    \label{eq:Hlm}
\end{equation}
where $\hbar\Delta/2$ is the energy gap between the two atomic levels, $\omega$ is the photon frequency and $\tilde{\lambda}_{1,2}/\hbar=\lambda_{1,2}$ are the interaction strengths between photons and atoms.
Such systems have been studied as the quantum Rabi model \cite{Rabi1936, Xie2017}, $H_{R}\!=\!H_{LM}(N\!=\!1,\lambda_1\!=\!\lambda_2\!=\!\lambda)$ for a single atom and the Dicke model \cite{Dicke1954,Kirton2018} $H_{D}\!=\!H_{LM}(\lambda_1\!=\!\lambda_2=\lambda)$ for $N$ atoms, which have $\mathbb{Z}_2$ symmetry with conserved parity $e^{i\pi(\hat{n}+\hat{J}_z)}$. In the absence of counter-rotating terms, i.e. $\lambda_1\!=\!\lambda$ and $\lambda_2\!=\!0$, there is $\text{U}(1)$ symmetry in the models, and they are termed the Jaynes-Cummings model \cite{Jaynes1963}, $H_{JC}\!=\!H_{LM}(N\!=\!1, \lambda_2\!=\!0)$ and the Tavis-Cummings model \cite{Tavis1968} $H_{TC}\!=\!H_{LM}(\lambda_2\!=\!0)$. These models have been experimentally realized with both weak and strong couplings. See Ref.~\cite{FornDiaz2019} for a recent review of light-matter interactions in the experimental setup.

\begin{figure}
    \centering    
    \includegraphics[width=1.0\linewidth]{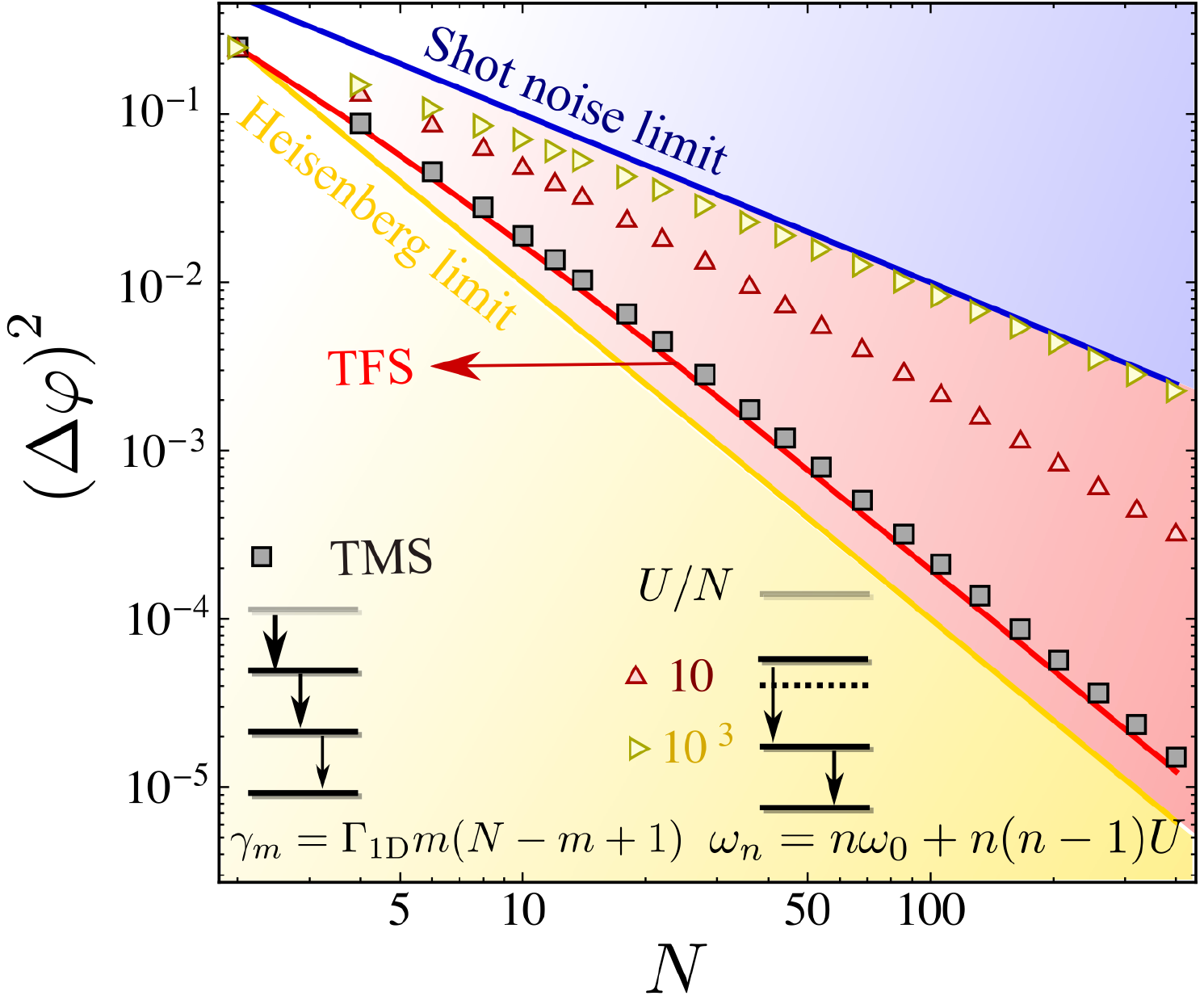}    \caption{\emph{\textbf{Scaling of sensitivity with system size of Dicke states.}} 
    {\color{black}{ The sensitivity of phase estimation $(\Delta\varphi)^2$ in an interferometric protocol with encoding on twin multimode states (TMS) is obtained using the Dicke model in superradiant phase and with parity measurements. Therefore, the Heisenberg limit is achievable here (black squares - twin multi-mode Dicke states), which reduces to the shot noise limit in presence of errors, which is anharmonic cavities  (upward red triangles corresponds to $U/N=10$, while downward yellow triangles corresponds to $U/N = 10^3$) in this case. The figure is taken from Ref.~\cite{Paulisch2019}.}} }
    \label{fig:supperrad_scal}
\end{figure}

These systems exhibit a phase transition at zero temperatures in the limits (a)  $N\to\infty$ and (b) $\Delta/\omega\to\infty$ with finite $\omega\Delta$ when $N=1$. While case (a), involving infinite atom number, is widely recognized as the thermodynamic limit,  the same terminology is used for case (b) as well by some authors. The phases can be distinguished by photon number in the ground state $\langle\hat{n}\rangle_{gs}$, with $\langle\hat{n}\rangle_{gs}$ independent of $N$ or $\Delta$ in the normal phase (NP), while above a critical interaction strength $\lambda^{cr}$, the underlying symmetry is broken and the ground state is in the superradiance phase (SRP), and the atoms coherently emit photons with $\langle\hat{n}\rangle_{gs}$ increasing with increasing $N$ or $\Delta$ exhibiting superradiance. The ground states are squeezed non-classical states~\cite{Ashhab2010, Gietka2023} and the critical interaction strengths are given by $\lambda^{cr}_{JC}=\sqrt{2\omega\Delta}$, and $\lambda^{cr}_{R}= \lambda^{cr}_{TC}=\lambda^{cr}_{D}=\sqrt{\omega\Delta}$ for the respective models. The symmetry of the model aids in the enhanced estimation of the symmetry-breaking field~\cite{Ivanov2013, Lorenzo2017} with HL achievable by measurement of magnetization. The superradiant photonic state of the Dicke model, which are twin multimode states, also shows HL in the interferometric setup~\cite{Paulisch2019} as shown in Fig.~\ref{fig:supperrad_scal}, when the states are used to encode a phase shift between two different paths of the interferometer.

The ground states of Dicke model $\hat{H}_D$ is endowed with quantum advantages, with both the spin degrees of freedom $A$ and the photonic degrees of freedom $B$~\cite{Wang2014}, achieving better precision than SQL. The parameter is encoded in the ground state $\rho$ of $\hat{H}_D$, either the photonic subsystem $\rho_A$ or the atomic subsystem $\rho_B$ via unitary evolution $\rho_\theta=e^{-i\theta G}\rho_{A,B} e^{i\theta G}$ with $G=\hat{b}^\dagger\hat{b}$ or $G=\hat{J}_x$ for photon or atoms, respectively. Although the QFI of global state, studied as fidelity susceptibility~\cite{Tsang2013}, achieves HL, the QFI $\mathcal{F}_\theta$ of each subsystem is discontinuous at critical points and scales super-linearly, scaling with both photon number and number of atoms, beating SQL near the superradiant phase transitions in the thermodynamic limit. Such an advantage arises from the increased squeezing of both the photons and the spins, which is reflected in the decreased value of the spin-squeezing parameter \(\xi_N^2\), in the ground state near the phase transition point. Moreover, the highest precision, given by QFI, can be accessed via homodyne measurements~\cite{Bina2016}. Interestingly,  light-matter systems $H_{LM}$ can gain quantum advantage in the metrological protocol, even if the encoding time, $\theta$, is explicitly taken into account. Various quench protocols has been investigated for $H_{LM}$. For the Rabi model, the ground state of the uncoupled $H_{R}$, i.e., $\lambda_{1,2}=0$, is given by $\ket{\psi_{0}}=\ket{0}_B\otimes\ket{\downarrow}_A$, where $\ket{\downarrow}_A$ is the ground state of the atom, and $\ket{0}_B$ is the photonic vacuum. Adiabatically quenching $\ket{\psi_{0}}$ in the NP, the ground state obtained can be represented as $\ket{\psi_{\lambda}}=\hat{\mathcal{S}}(\xi)\ket{0}_B\otimes\ket{\downarrow}_A$, where $\hat{\mathcal{S}}(\xi) = \exp [\frac{\xi(\hat{a}^{\dagger})^{2}-\xi^{*} \hat{a}^{2}}{2}]$ is the squeezing operator with $\xi=\frac{1}{4}\log\left[1-g^2\right]$ and $g=\lambda/\lambda^{cr}_R$ \cite{Gietka2023}. The QFI of $\ket{\psi_{\lambda}}$, while sensing the parameter $\theta=\omega$, is given by $\mathcal{F}_\theta\sim\left[\theta\left(1-g\right)\right]^{-2}$~\cite{Gietka2022a, Garbe2022, Gietka2022b}, which diverges near criticality, $\lambda\to\lambda^{cr}$, i.e., $g\to 1$. With the finite time ramping in interaction strength, such as $\frac{dg}{dt}=\gamma\omega(1-g^2)^{2/3}$, where $\gamma\ll 1$, the time to reach from  $g=0$ to $g\lesssim1$ is given by $T\sim (\gamma\omega)^{-1}(1-g^2)^{-1/2}$, which indeed diverges for as $g\to1$, where $T$ represents time of the quench. Using such time-dependent situations, it has been shown in the Rabi model $H_R$, that $\mathcal{F}_\Delta\sim T^4$~\cite{Garbe2020, Gietka2021}, i.e., gives the Heisenberg limit, which reduces to the SQL $\mathcal{F}^{diss}_\Delta\sim T^2$ in the presence of dissipation. A sub-HL scaling is achieved by Fisher information of photon statistics in the limit of continuous measurement~\cite{Ilias2022}. In the sudden quench protocol with the Rabi model, it has been shown that for a specific class of the initial states, when evolved with the critical Hamiltonian, $H_R(\lambda=\lambda_R^{cr})$, homodyne measurements can lead to saturation of the QFI~\cite{Chu2021}. Generalizing between the adiabatic and the sudden quench protocol through finite time protocols, the bounds for each evolution are examined ~\cite{Garbe2022}, which reports scaling, $\mathcal{F}_\theta\sim T^4$, achievable via homodyne measurements. The sudden quench protocol gives $\mathcal{F}_\theta\sim T^6$, a super-HL scaling, attainable by non-standard measurements. While higher $T$ scaling can be obtained in transient times with small QFI, the studies reveal the highest QFI with the sudden quench protocol. Interestingly, exponential scaling of $\mathcal{F}_\omega$ with $T$ is demonstrated with non-adiabatic quenching towards the triple critical points, ($\lambda_1\to\lambda_{R}^{cr}, \lambda_2\to0$), in the anisotropic Rabi model with both rotating and counter-rotating terms. This is, however, suppressed in the presence of dissipation~\cite{Cheng2025}.

Further analysis of the sudden quenching in the NP of the Dicke model has been performed~\cite{Gietka2022b, Gietka2022c}. Particularly, exponential scaling $\mathcal{F}_\theta\sim\exp(T)$ is obtained while quenching in the SRP. Adiabatic quenching of the Dicke model in the NP, but with displaced initial states, HL is obtained for $\mathcal{F}_\Delta\sim T^4$, thereby giving enhanced precision. Dynamical states generated by Tavis-Cummings from various initial states, namely Dicke states, GHZ states, and X-polarized states of qubits with photon vacuum, can enhance the sensing of the coupling parameter $\lambda_1$ with HL in the presence of resonator and qubit decays. At large decay rates, X-polarized states provide maximum QFI~\cite{Saleem2024}. The  adiabatic quenching of a single squeezed photonic state together with the atomic ground state of $H_{TC}$ yields spin-squeezed states, which can facilitate super-HL scaling in interferometric sensing for large squeezing parameters~\cite{Pavlov2025}.
Sudden quenching with the Jaynes-Cummings model can produce the two-mode squeezed states, which are useful for quantum sensing~\cite{Cardoso2021, Lu2024}. In the presence of non-linear photonic effects, given by $(\hat{b}^\dagger)^2+\hat{b}^2$ terms, the critical point of $H_{JC}$ shifts and the homodyne measurements give enhanced sensing near the critical points, even at finite $\Delta/\omega$~\cite{Lu2022}.  The nonlinear terms in the light-matter systems can lead to nonanalytic behaviour. These systems exhibit much higher measurement precisions due to their first-order-like phase transition~\cite{Candeloro2021, Ying2022}.


\subsection{Bose-Einstein condensates}
The Bose-Einstein condensate (BEC) ~\cite{Pitaevskii2016} has found extensive applications in quantum sensing, e.g., magnetometry, gravimetry, inertial sensors, etc ~\cite{Gross2012, Huang2014, Pezze2018, Szigeti2021, Huang2024}. Here we briefly discuss metrologically useful states and various methods to prepare them both in the context of interferometry-based sensing and criticality-based sensing.  
\subsubsection{Metrologically-useful states}
\label{metro_BEC}
\textbf{\emph{Spin squeezed states.}} Quantum sensing of unknown parameter often involves interferometry~\cite{Torii2000,Cronin2009,Ockeloen2013,Seifert2025} of quantum states where the parameter is encoded as the phase difference between the two arms of the interferometer. Various interferometric protocols~\cite{Schumm2005, Sadgrove2013, Marti2015, Gebbe2021, LeDesma2025, Bell2016} are undertaken for sensing purposes. Their operating principle is based on the interference of two coherent atomic modes. In such sensors, the quantity to be estimated is mapped to the relative phase between the two modes given by the two annihilation operators $a$ and $b$, respectively. These two bosonic modes can be written in terms of effective spin operators defined as, 
    $\hat J_+= b^\dagger a, \quad \hat J_- = a^\dagger b, \quad \hat J_x = \frac{1}{2}\left( \hat J_+ +\hat J_-\right),
     \hat J_y = \frac{1}{2i}\left( \hat J_+ - \hat J_-\right), 
     \quad  \text{and} \quad \hat J_z = \frac{1}{2}\left(n_b - n_a\right)$,
where $n_a = a^\dagger a$ and $n_b = b^\dagger b$, with $a^\dagger$ and $b^\dagger$ being the creation operators of mode $a$ and mode $b$, respectively.

Any coherent spin state $|\alpha, \varphi\rangle$ can be constructed by unitarily rotating the state $|0,0\rangle = \frac{1}{\sqrt{N!}}(b^\dagger)^N|\text{vac}\rangle$, $N$ being the number of bosons in the mode $b$. The coherent spin-state is given by,
    $|\alpha, \varphi\rangle = \frac{1}{\sqrt{N!}} \left[ \sin({\alpha}/{2}) e^{-i\varphi/2}a^\dagger + \cos({\alpha}/{2}) e^{i\varphi/2}b^\dagger  \right]^N|\text{vac}\rangle$
for a $N$-bosons. Note that there is no quantum correlation among the bosons. As discussed in Sec.~\ref{interferometry_squeezing}, spin-coherent states do not offer any quantum advantage in metrology. However, it is possible to improve the sensing capability of the state by employing spin-squeezing \cite{Gross2010,Lucke2011,Liu2011,Hamley2012,Luo2017,Zou2018,Mao2023}. Here we discuss some ways in BEC systems to produce spin-squeezed states.

\par
\textbf{\emph{Bosonic Josephson junction.}} The two-mode BECs consist of atom-atom interactions. These two modes of the BEC can either be due to the external trapping potential or internal degrees of freedom. The two-mode BEC is modelled using the Bosonic Josephson junction (BJJ) Hamiltonian 
\begin{align}
    {H}_{\text{BJJ}} = -\hbar \Omega \hat{J}_x + \hbar \chi \hat{J}_z^2 + \hbar \nu \hat{J}_z.
\end{align}
As described earlier, the operators $\hat{J}_z$ and $\hat{J}_x$ are constructed using the operators of the two modes. The first two terms with parameters $\chi$ and $\Omega$ simulate the LMG model ($\gamma=0$) and are control parameters. The ground state of the Hamiltonian ${H}_{\text{BJJ}}$ possesses quantum entanglement that is useful for quantum metrology~\cite{Gross2012, Pezze2018}.
 Now let us consider the case where the differential energy shift $\nu$ is set to zero. It is known that this system has three regimes based on the relative strength of tunnelling coupling, $\chi$ which is given by $\Lambda = N\chi/\Omega$. The three regimes are as follows: (a) Rabi regime, where $\Lambda \ll 1$ in the ground state is a coherent spin state, where $\langle \hat J_x\rangle  = N/2$. The QFI scales as $\mathcal{F} = N\sqrt{\Lambda +1}$. (b)  Josephson regime, where $1<\Lambda<N$, with the ground state being a coherent spin squeezed state with reduced spin fluctuation in $\hat J_z$ compensated by increased fluctuations in $\hat J_y$. The ground state is therefore suitable for quantum-enhanced sensing and QFI $\mathcal{F} \sim N^{3/2}$. (c) Fock regime, where $\Lambda \gg N$ and  the ground state is a twin-Fock state that is an eigenstate of $\hat J_z$ with eigenvalue zero. The QFI attains HL scaling with system size. For negative values of $\Lambda$, there are two phases--disordered $(-1<\Lambda <0)$ and ordered $\Lambda<-1$. The QFI gives the HL scaling with system-size in the ordered phase. Recently, the authors in Ref.~\cite{Sorelli2019} have investigated the role of $\Lambda$ in the dynamics of generation of metrologically useful entanglement. They found that the dynamical generation of entanglement is fastest when the coherent spin state points along the negative x-axis and
$\Lambda =2$ and that the linear coupling accelerates the dynamical creation of entanglement when compared to one-axis twisting dynamics. The enhancement of entanglement has also been considered using oscillatory dynamics of squeezing~\cite{Zhang2024}, using chaotic dynamics in a periodically driven BJJ~\cite{Liu2021}, spin squeezing in a spin-1/2 BEC with spin-orbit coupling~\cite{Chen2020}, spin-nematic squeezing~\cite{Mao2023} and simultaneous spin-momentum squeezing~\cite{Wilson2022}. There are several strategies to generate coherent spin squeezed states, which can be broadly classified according to the adiabatic and the diabatic approaches. In the adiabatic approach, $\Lambda$ is changed slowly, whereas in the diabatic regime, the parameter $\Lambda$ is changed fast. Recently, shortcut to adiabaticity protocols have been utilized to generate entangled states in BJJ in Refs.~\cite{Hatomura2018, Odelli2023}. Certain dynamical methods, such as one-axis twisting (OAT), two-axis counter-twisting~\cite{Kitagawa1993} and twist-and-turn dynamic, are the usual diabatic methods used to generate spin squeezed states~\cite{Davis2016, Frowis2016, Nolan2017}.

In OAT, the system is evolved using $e^{-i H_{\text{OAT}}t/\hbar}$, where $ H_{\text{OAT}} = \hbar \chi \hat J_z^2 $. This can be viewed as a $\langle \hat J_z\rangle$-dependent rotation about the $z$-direction, leading to twisting of the state on the Bloch sphere with squeezing reading to an optimal value $\xi \sim N^{-1/3}$ at optimal time $t \sim N^{-2/3}/\chi$. It is possible to produce a highly spin-squeezed state using two-axis counter-twisting (TACT) as well. Here $ H_\text{TACT} = \hbar\chi (\hat J_x^2 - \hat J_y^2)$ with squeezing $\xi \sim N^{-1}$.
The experimental progress of TACT is hindered due to the presence of unwanted noise in the detection of the output quantum state. To overcome this, spin echo protocol has been proposed~\cite{Anders2018, Ma2024} with TACT.  Another commonly used technique for achieving squeezing is twist-and-turn (TNT) dynamics~\cite{Muessel2015} with Hamiltonian $ H_\text{TNT} = \hbar \chi \hat J_z^2 + \hbar \Omega \hat J_x$. Recently, TNT dynamics in BJJ is enhanced with shortcut to adiabaticity protocol~\cite{Odelli2024} and machine optimization~\cite{Huang2022,Zhou2024}.

\textbf{\emph{Spinor Bose-Einstein condensates.}} The constituent particles of BEC can have certain spin degrees of freedom in addition to the spatial degree of freedom. A spin-$F$ boson can be described using a $2F+1$-dimensional vector. Collisions in BECs can lead to spin exchange of the particles~\cite{Law1998,Pu1999}. Under the single mode approximation, it is assumed that the spin dynamics do not affect the spatial distribution of the condensate. For a spin-$1$ BEC, the spin exchanging Hamiltonian is given by $ H_\text{SM} = (c/2N)\hat S^2 - qN_0$~\cite{Zhang2013}, where $c$ and $q$ are the inter-spin and effective quadratic Zeeman energies, respectively. The spin operators $\hat S^2 = \sum_i \hat S_i^2$, where $\hat S_i = \hat a^\dagger _\alpha S^{\alpha\beta}_i \hat a_\beta $, with the operators $\hat a_\alpha$ and $\hat a_\alpha^\dagger$ being the creation and annihilation operators, respectively, for the spin modes $\alpha = \pm1$ and $0$. The ground state of the spinor-BEC model, is metrologically useful. It has been demonstrated in experiments~\cite{Klempt2010,Scherer2010} that the parametric amplification of vacuum fluctuations in the generation of $Rb$ atoms ($F=2$) can help to generate squeezing and entangled states.  Recently, a protocol to generate macroscopic superposition states of a large number in the ground state of a spin-1 BEC has been proposed in Ref.~\cite{Pezze2019b}.  These states are prepared with around $90\%$ probability by measuring the number of particles in one of the three modes. The protocol has been shown to be robust and implementable with current experimental capabilities. In Ref.~\cite{Zhou2023}, the spin-1 BEC initially prepared in a coherent spin state can be used to probe the parameter $q$ in $ H_\text{SM}$ with sensitivity beyond SQL.



\textbf{\emph{Critical sensing in BEC.}} Quantum criticality has been extensively used as a resource for sensing. For BECs, critical-sensors have been proposed in a few works~\cite{Pezze2019,Mirkhalaf2020,Mirkhalaf2021}. In Ref.~\cite{Mirkhalaf2020}, two different critical points were investigated for quantum-enhanced sensing, specifically, transitions from polar to  broken-axisymmetry and from antiferromagnetic to broken-axisymmetry phases for a spin-1 system given by the Hamiltonian $H_\text{SM}$, with $q$ being the parameter sensed. The sensitivity was determined using the QFI and the precision scales with the number of atoms up to $N^4$ around the criticality. Recently, in Ref.~\cite{Debnath2025}, it has been illustrated that a BEC under a tilted external potential can act as a probe for quantum metrology. The BEC is under the external potential $V_\text{ext}(x) = E\sin^2(kx) + f|x|$, the quantity $\tilde V \coloneq f/E$ is varied to observe a delocalisation-to-localisation transition as $\tilde V$ is increased. The QFI, when estimating $\tilde V$, has a super-HL scaling which despite decreasing with an increase in non-linear interaction remains super-HL for a wide range of values of non-linear interaction.
\subsubsection{Applications}
\begin{figure*}
    \centering
    \includegraphics[width=\linewidth]{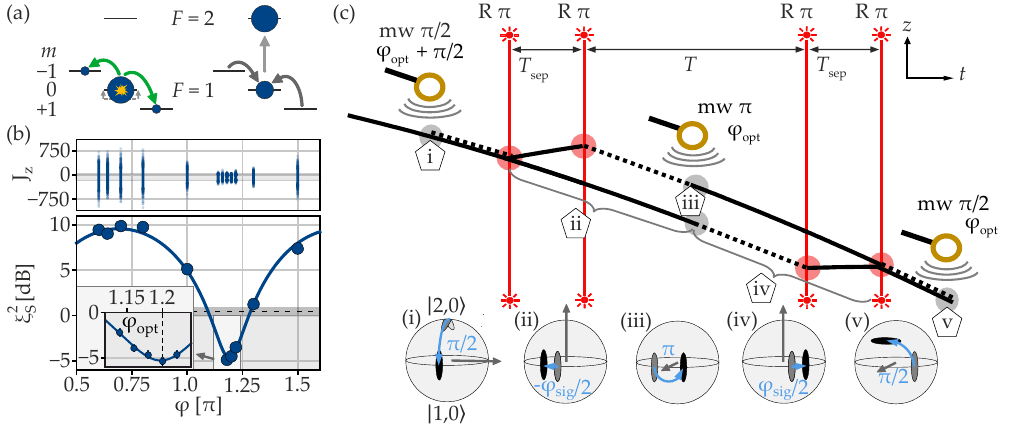}
    \caption{\textbf{\emph{Gravimeter based on Bose-Einstein Condensate.}} 
    (a) Schematic depicting the activation of spin-collisions dynamics (green arrows) between atoms by dressing the transition $|1,0\rangle \leftrightarrow |2,0\rangle$ with mw pulses leading to the generation of two-mode squeezed vacuum state, transfer of $|1,0\rangle$ to $|2,0\rangle$ using mw (light gray arrows) and rf pulse induced transfer from $|1,\pm1\rangle$ to $|1,0\rangle$ (dark gray arrows).  
    (b) Spin noise tomography of the input state of the interferometer is presented. The normalized population in $|2,0\rangle$ and the corresponding variances are presented against the scanned mw phase $\varphi$ in the top and bottom graphs respectively, with the inset focusing on the around the minimum that represents the optimal squeezing angle $\varphi_\text{opt}$. 
    (c) Schematic depiction of the whole sequence of operation for the gravimeter. The dashed line indicate $|1,0\rangle$ and solid line $|2,0\rangle$. The Raman (R) pulses are depicted in red. In the bottom figure, the bloch sphere is depicted with $|2,0\rangle$ and $|1,0\rangle$ being the north and south pole respectively. The squeezed input state is rotated into the phase-squeezed direction (i) and the phase $\varphi_\text{sig}$ that is to be sensed is encoded in (ii)-(iv), which is finally estimated by measuring population imbalance in (v).  
    The figure is taken from Ref.~\cite{Cassens2025}.}
    \label{fig:gravimetry}
\end{figure*}
    \textbf{\emph{Thermometry.}}
    Thermometry is one of the most widely used applications of quantum metrology. In quantum mechanics, temperature is not an observable and thus cannot be directly measured. However, temperature can be deduced indirectly from the measurements of observables of a quantum system. This is the goal of thermometry; for a review, see Ref.~\cite{Mehboudi2019b}. BEC-s are the most commonly used platforms for thermometry of very low temperatures. Thermometry in  BEC-s is roughly based on two techniques, introducing impurity in the BEC~\cite{Sabin2014, Olf2015, Mehboudi2019b} and time-of flight measurements~\cite{Leanhardt2003, Gati2006a, Gati2006b}. The impurity-based techniques are used to measure sub-nano-kelvin temperatures efficiently, but they destroy the BEC state due to backaction. In~\cite{Mehboudi2019a}, the authors realize non-demolition temperature measurement by considering the Bose-polaron model. The global Hamiltonian is as follows:
    \begin{align*}
        {H}_{IB} = \frac{\hat p^2}{2m_I} + \frac{m_I \Omega^2}{2}\hat x^2 + \sum_k E_k \hat b_k^\dagger \hat b_k + \sum_k \hbar g_k \hat x(\hat b_k + \hat b_k^\dagger).
    \end{align*}
    The first two terms corresponds to the Hamiltonian of the impurity trapped in a harmonic potential of frequency $\Omega$. The third term is the Hamiltonian of the BEC, while the last term is the interaction between the impurity and the BEC. The impurity is modelled as a Brownian particle, and its steady-state variances, $\langle \hat x^2 \rangle$ and $\langle \hat p^2\rangle$ have the temperature, $T$, of the BEC encoded in them. The SLD and QFI for temperature turn out to be
    \begin{align*}
        \hat \Lambda_T &= C_x(\hat x^2 - \langle \hat x^2 \rangle ) + C_p(\hat p^2 - \langle \hat p^2 \rangle), \quad \text{and}\\
        \mathcal{F}_T &=  2C_x^2\langle \hat x^2\rangle^2 + 2C_p^2\langle \hat p^2\rangle^2 - \hbar^2 C_xC_p \quad 
    \end{align*}
    where $C_x = ({4\langle \hat p^2\rangle^2 \chi_T(\hat x^2) + \hbar \chi_T(\hat p^2)})/(8\langle \hat x^2\rangle^2\langle \hat p^2\rangle^2 - \hbar^4/2)$, with $\chi_T(\hat O) \coloneqq \frac{1}{2} \langle \hat O \hat \Lambda_T +  \hat \Lambda_T \hat O \rangle - \langle \hat O \rangle \langle \hat \Lambda_T \rangle$, and $C_p$ is obtained by interchanging $\hat x$ and $\hat p$ in $C_x$. The technique introduced outperforms the usual thermometric technique, e.g., see~\cite{Sabin2014}, by an order of magnitude and is nondestructive. In addition to these techniques, it has been shown in~\cite{Planella2022} that bath-induced correlations can enhance thermometry. In this technique, multiple probes that are not interacting among themselves interact with cold bosonic baths. This may lead the probes to develop correlations among themselves in the steady state.  Criticality-based thermometry has also been studied in the literature using BECs. For example, in Ref.~\cite{Aybar2022}, the authors studied thermometry with finite-sized strongly correlated systems exhibiting quantum phase transition in the spin-1 spinor BEC model given by $ H_\text{SM}$ (see Sec.~\ref{metro_BEC}). The probe state is considered to be the Gibbs state $\hat \rho(T,\lambda) = \sum_n \frac{e^{-E_n/T}}{Z}|\psi_n\rangle\langle\psi_n|$, consequently the QFI $\mathcal{F}_T(T,\lambda) = \delta^2\hat H(T,\lambda)/T^4$. The ground state of the system consists of two critical points, one that separates the Polar phase from the broken axisymmetry (BA) phase and another that separates BA from the antiferromagnetic (AFM) phase. They demonstrate the scaling $\mathcal{F}_T \sim N^{2/3}$ at both the critical points. They also show that the system has a first-order phase transition in the AFM condensate for the zero magnetization case,  and the QFI demonstrates $\mathcal{F}_T \sim N^2$.
    
    \textbf{\emph{Inertial sensors and gravimeters.}} Cold atoms platforms, especially BEC-s are extensively used as inertial sensors~\cite{Kapale2005, Pandey2019, Stolzenberg2025}, gyroscopes~\cite{Cooper2010, Gutierrez2015, Cooling2021, Skulte2023}, gravimetry~\cite{Debs2011,Zhou2012, Mutinga2013, Carraz2014, Aguilera2014, Condon2019, Trimeche2019,  Szigeti2020, Ufrecht2021, Cassens2025}, and detection of gravitational waves~\cite{Howl2023,Yu2024,Sen2024,Schutzhold2018,Robbins2019,Sabin2016, Sabin2014b}. Refs.~\cite{Robins2013,Meister2017, Geiger2020} review these topics. These sensors are mostly based on atom interferometry~\cite{Baudon1999,Hogan2009,Abend2019}, where atom clouds are made to free-fall and laser-pulses are used to coherently split the ensemble into two, which after the free-fall, are recombined using light pulses to form an interference pattern. In Ref.~\cite{Cassens2025}, the authors present a gravimetry protocol (see Fig.~\ref{fig:gravimetry}), where a sensitivity of $-1.7^{+0.4}_{-0.5}$dB beyond the SQL is demonstrated. Initially a BEC of $6\times10^3$ ${}^{87}$Rb atoms are prepared in a dipole trap with spin level $|F,m\rangle = |1,0\rangle$. A homogeneous magnetic field orientated in parallel to the Earth's gravity is applied to prevent spin-changing collision and suppress the creation of $|1,\pm1\rangle$ states. A blue-detuned microwave (MW) field is used for $50$ ms to activate the clock transition $|1,0\rangle \leftrightarrow |2,0\rangle$, this populates the levels $|1,\pm1\rangle$ with a two-mode squeezed vacuum state.  The dipole trap is switched off after this for 1 ms and the system undergoes free fall, following which the dipole trap is turned on again for 350 $\mu$s in order to slow down the expansion of the cloud. Atoms in $|1,0\rangle$ are transferred to $|2,0\rangle$ by application of a $\pi$ pulse. Atoms in $|1,\pm1\rangle$ are then transferred to $|1, 0\rangle$ by a $\sigma^-$-polarized radio-frequency (RF) $\pi$ pulse with phase $\phi_{\text{rf}}$. After this the ensemble is placed in the interferometer. The single-mode squeezed vacuum state  in $|1,0\rangle$ and the state $|2,0\rangle$ forms a spin-squeezed state which leads to entanglement-enhanced gravimeter. First an MW $\pi/2$ pulse with a phase $\varphi_{\text{opt}} + \pi/2$ is applied. After 1.9 ms, a Raman $\pi$ pulse driving transition $|1,0;p=0\rangle\rightarrow |2,0;p=\hbar k_{\text{eff}}\rangle$  is applied, which leads to the spatial delocalization of the two momentum modes. The two clouds separate for $T_{\text{sep}} = 77 \mu$s  before a second Raman $\pi$ pulse decelerates the upper arm of the interferometer by driving the same transition.
    While both atomic clouds fall with the same momentum for a certain duration, a resonant mw $\pi$-pulse is applied to invert their internal spin states, acting as a spin-echo to cancel undesired spin evolution. After this, the two interferometer arms acquire an additional gravitational phase shift of opposite sign, which is not canceled by the echo and thus encodes the phase that is to be estimated. The wave packets are then recombined by applying identical Raman processes to the lower arm of the interferometer. After $1.9$ ms, the microwave $\pi$-pulse the imprinted inertial phase with squeezed quantum noise, is mapped onto the population imbalance $|2,0\rangle$ and $|1,0\rangle$ states by a MW $\pi/2$ pulse with phase $\varphi_{\text{opt}}$, leading to estimation of the gravitational constant $g$.

\textbf{\emph{Magnetometry.}} Magnetometers are used to probe magnetic fields in physical systems~\cite{Greenberg1998,Tsuei2000}, biomedical systems~\cite{Hamalainen1993}, etc. Recently, magnetometers on BEC-platform~\cite{Gunther2005, Wildermuth2005, Wildermuth2006, Vengalattore2007, Aigner2008, Ng2013, Steinke2013, Eto2013,Muessel2014, Jasperse2017, Yang2017, Apellaniz2018,  Czajkowski2019, Yang2020, PalaciosAlvarez2022, Shukla2024} have been extensively studied. In Ref.~\cite{Vengalattore2007}, a system of spinor BEC of $^{87}$Rb atoms has been used to demonstrate magnetometry.  The spin-polarized spinor BEC atoms are prepared in an optical trap. The local magnetic field to be estimated is encoded in the Larmor precision of the spins. To do this, a RF  pulse is used to tip the magnetization perpendicular to a bias field imposed along the axis of the condensate. The spins precess at a rate proportional to the local magnetic field to be estimated. The condensate is probed using magnetization-sensitive imaging to detect the spin orientations. These detections help in assessing the spin precession rate and thereby the magnetic field. In contrast to this, continuous Faraday measurements are also employed to detect the spin orientations in various magnetometers like in Ref.~\cite{Jasperse2017}. The field sensitivity over a measurement area $A$ comes out to be $\delta B = (\hbar/g \mu_B)(1/\sqrt{\tau D T})(1/\sqrt{\tilde n A})$, where $\tau$ is the Zeeman coherence time and $\tilde n$ is the local column density of the gas. The measurements are repeated over total time $T$ at a duty cycle $D$. The protocol demonstrates improvements over other magnetometers like SQUID (magnetometers based on superconducting Josephson junctions~\cite{Kirtley1995, Bending1999}) especially for low-frequency fields. In Ref.~\cite{Yang2017}, inhomogeneous magnetic fields are estimated with the BEC. The inhomogeneous magnetic fields exert Zeeman force on the atoms trapped in a smoothly varying potential. The atoms move under this force, distorting the BEC wavefunction. The distorted BEC density is imaged  by recording the absorption of resonant light using a CCD camera, thereby mapping the magnetic field.

\textbf{\emph{Clocks.}} Atomic clocks are high-precision timekeeping devices that have been implemented on atomic ensembles~\cite{Martin2013, Andre2004, Borregaard2013, Kruse2016}. The basic principle behind them is to estimate the frequency of radiation that the atoms absorb or radiate to transition between two atomic levels. Once the frequency is fixed, the cycles can be counted and the time can be measured. In Ref.~\cite{Kruse2016}, an atom interferometer in clock configuration is designed and implemented. In the proposed clock, $N = 10^4$ atoms are combined in one input state with a quadrature squeezed vacuum with an average of $0.75$ atoms in the second input. Usually in interferometry, there are two inputs to the interferometer, where an ensemble of atoms is prepared in one of the inputs, while the other one is left empty (vacuum). The clock surpasses the SQL by squeezing vacuum in the usually empty input. The squeezed vacuum is generated by spin-changing collisions in a BEC of $^{87}$Rb atoms. The clock consists of a four-modes given by the states, $|\pm 1\rangle = |F = 1, m_F = \pm 1\rangle$, $|0\rangle = |F = 1, m_F = 0\rangle$, and $|e\rangle = |F = 2, m_F = 0 \rangle$, which is reduced to an effective two-level system in terms of the states $|g\rangle = (|+1\rangle + |-1\rangle)/\sqrt2$ and $|e\rangle$.  The interferometer effectively only couples the $|g\rangle$ and $|e\rangle$, without affecting the antisymmetric state $|h\rangle = (|+1\rangle - |-1\rangle)/\sqrt 2$. The atoms in the $|e\rangle$ state picks up a phase as they evolve in time. This phase is estimated from the occupation of the three Zeeman states, with a sensitivity of $2.05$dB below the SQL.    

\section{Sensing in lattice systems}
\label{sec-lattice}

This section describes systems of one or more particles arranged on a lattice, where the lattice points are spatially separated and structured in a regular repeating pattern, typically in one- ($1$D), two- ($2$D) and three-dimensions ($3$D). These lattice models can offer a platform to simulate and understand a wide range of microscopic phenomena in condensed matter systems and to build quantum devices in a controlled manner. In our context, particularly, we focus on the role of these models in quantum sensing which involves the detection of several parameters like external magnetic and electric fields, amplitude of hopping, etc.

Among them, the most prominent examples include the Bose-Hubbard (BH) and Fermi-Hubbard (FH) models, which describe interacting bosons and fermions on a lattice, respectively, nearest-neighbor interacting quantum spin models, like Ising, Heisenberg models \cite{vandar1993}, Su–Schrieffer–Heeger \cite{Su1979} and Kitaev model \cite{Kitaev2001} with long-range interactions. These models serve as paradigms for studying various quantum phases, such as superfluidity, Mott insulators, ferro- and paramagnetism and superconductivity, that can be exploited to achieve enhanced sensing. Interestingly, it was found that there are lattice without interactions such as the Aubry-Andr{\'e} (AA) model \cite{Sahoo2024a} and the system with Stark onsite potential \cite{He2023} showing localization-delocalization transitions due to quasi-periodic or linear potentials, respectively can provide Heisenberg scaling in QFI. Further, non-Hermitian models, like, the Hatano-Nelson (HN) model with asymmetric hopping \cite{Edvardsson2022, Gohsrich2025}, further enrich this landscape by introducing phenomena like exceptional points and the non-Hermitian skin effect, which have also been shown to increase sensing precision. 


 \subsection{The Fermi-Hubbard model}

The Hamiltonian of the Fermi-Hubbard model can be described as 
\begin{equation}
\label{eq:FH}
H_{\text{FH}} = -t \sum_{\substack{\langle i,j \rangle\\ \sigma=\uparrow,\downarrow}} \left( c^\dagger_{i\sigma} c_{j\sigma} + \text{H.c.} \right) + U \sum_{i} n_{i\uparrow} n_{i\downarrow} - \mu \sum_{i,\sigma} n_{i\sigma},
\end{equation}
where $c_{i\sigma}^\dagger$ and $c_{i\sigma}$ are the fermionic creation and annihilation operators for an electron with spin $\sigma$ (either $\uparrow$ or $\downarrow$) at lattice site $i$. The parameter $t$ represents the hopping amplitude between nearest-neighbor sites $i$ and $j$ using the notation $\langle i, j \rangle$, $U$ denotes the onsite repulsion interaction strength between two electrons occupying the same site, and $\mu$ is the chemical potential. At zero temperature ($T = 0$), the FH model exhibits different phases depending on the filling factor (average number of particles per site), hopping amplitude $t$, and the interaction strength $U$ \cite{Essler2005}. When the interactions are weak compared to the hopping strength ($t \gg U$) and the lattice is less than half-filled (i.e., fewer than one fermion per site), the electrons can move freely throughout the lattice. This phase is known as the metallic phase, which allows for electrical conduction. But, when the interaction strength dominates (i.e., $t \ll U$) and the system is at half-filling, meaning there is exactly one particle per site, the system behaves as an insulator, known as Mott insulting phase. This is because particles cannot easily move to neighboring sites due to the strong repulsive interaction $U$, which makes it energetically costly to place an additional particle on an already occupied site, and hence particles become localized. However, their spins still interact with each other, resulting in a type of magnetic order, called antiferromagnets, where neighboring spins align in opposite directions ($\uparrow\downarrow\uparrow\downarrow \dots$). By doping the Mott-insulating phase, i.e., by introducing holes or electrons into the system, the magnetic order gets destroyed and there is an evidence that the superconducting pairs, characterized by d-wave symmetry are formed \cite{Scalapino1995, White1998}.
 
An experimental milestone was achieved by observing the Mott insulating phase using ultracold fermionic potassium-40 ($^{40}\mathrm{K}$) atoms with single-site resolution in a two-dimensional optical lattice \cite{Cheuk2016}. Similarly, direct observation of antiferromagnetic correlations in the FH model were reported with ultracold lithium-6 ($^{6}\mathrm{Li}$) atoms cooled to near absolute zero and confined in a $1$D and $2$D optical lattice \cite{Parsons2016, Boll2016, Mazurenko2017}.

 \subsubsection{Sensing in interacting fermionic model} 

Multipartite entanglement is a crucial and powerful resource for achieving HL in the quantum-enhanced metrology although its detection in experiment is challenging. In this respect, entanglement criteria for systems having spin degrees of freedom have been developed in terms of QFI. Beyond spin systems, bounds connecting entanglement between modes and QFI are also obtained~\cite{deAlmeida2021}. Specifically, starting from a thermal state of the 1D Fermi-Hubbard model at short time scales, dynamical observables are shown to relate QFI and multi-mode entanglement. Further, the connection of QFI with Van Hove correlations, which is a real-space two-time correlation function accessible via neutron scattering experiment, is established for FH chain~\cite{Laurell2022}. In the pseudogap regime of the FH model, high entanglement, estimated via QFI, can be generated, and a lower bound on entanglement can be obtained using the dynamical vertex approximation. These theoretical results turn out to be consistent with experimental findings from neutron scattering experiments~\cite{bippus2025}. Several other studies have also explored the quantification of multipartite entanglement using QFI in the FH model~\cite{Tomasi2019, Sajna2020, kasatkin2024, kaczmarek2023, mirani2024}. 
On the other hand, Ref.~\cite{Mamaev2020} proposes a systematic method that flips one site at a time, for preparing generalized GHZ states using ultracold fermions in $3$D optical lattices or optical tweezer arrays. By exploiting the GHZ states and laser detuning, a parity measurement can ensure the higher precision of the relative phase, surpassing SQL~\cite{Hauke2016}. 

{\emph{\textbf{Spin squeezing in FH.} } }  In recent years, optical lattice clocks (OLCs) have emerged as one of the most impressive developments in precision measurement, achieving unprecedented levels of accuracy and stability. In particular, the sensitivity of atomic clocks can be improved~\cite{Sorensen1999} by using spin squeezing techniques like one-axis twisting and two-axis counter-twisting, which can be simulated from the FH model~\cite{Hern2022}. However this method has some flaws since creating entanglement requires atom-atom interactions, which decrease coherence in atoms required for atomic clocks.
Interestingly, recent work has shown that this drawback can be overcome by exploiting weak spin-orbit coupling (SOC) and minimizing deterioration of inter-atomic coherence. By coupling an external laser with the internal states of the atom, it is possible to generate SOC which can be described by the FH model. Here one of the hopping terms in Eq.~\ref{eq:FH} gets modified as $ e^{i\phi} c^\dagger_{j,\uparrow} c_{j+1,\uparrow}$ with $\phi=k_L a$ representing the SOC angle, $k_L$ being wave number of the laser and $a$ being the lattice-spacing.
It has been shown that SOC in this system gives rise to effective OAT dynamics, described by the Hamiltonian, $\frac{H_{\text{eff}}}{\hbar} = -\frac{U}{L} \, \mathbf{S} \cdot \mathbf{S} - \bar{B} S_z + \chi S_z^2$, where $\mathbf{S}$ is the collective spin operator and the nonlinear coefficient $\chi = \frac{\tilde{B}^2}{(N - 1) f U}$, with $\bar{B} = \frac{1}{N} \sum_q B_q$ being the mean axial field, $\tilde{B}^2 = \frac{1}{N} \sum_q (B_q - \bar{B})^2$ quantifying its variance~\cite{He2019} and $f$ being the filling fraction. This OAT squeezing procedure turns out to scale as $\xi \propto N^{-1/3}$ where $\xi$ is the Ramsey squeezing parameter, a performance quantifier for metrology. By applying an amplitude-modulated drive of the form $\frac{H_{\text{drive}}(t)}{\hbar} = \Omega_0 \cos(\omega t) S_x$, or continuous drive, the OAT system can be transformed to an effective two-axis twisting (TAT) one, which can provide scaling as $\xi \propto N^{-1/2}$, approaching the HL.


\emph{\textbf{Temperature sensing in FH. } }  Recent advancements in thermometry have also been successfully demonstrated in the FH model~\cite{Shen2023, Hartke2020}. Specifically, by using recursive algorithm, it was shown that the estimation of temperature can be made more accurate by considering canonical ensemble than the grand canonical ones, since the grand canonical ensemble converges to the ground state slower than the canonical ensemble in the presence of a finite number of particles. Extending $\text{SU}(2)$ $2$D FH model to $\text{SU}(N)$ ones (with $\sigma=\{1,\dots,N\}$) having higher symmetric group realizable via $^{173}\mathrm{Yb}$ or $^{87}\mathrm{Sr}$, temperature and entropy can be shown to be estimated~\cite{Wu2003, Pasqualetti2024}.


\subsubsection{Sensing in free fermionic model}
\label{critical_sensing_in_lattice}

It is extremely important to identify characteristics of many-body systems which are responsible for criticality-based sensing. It turns out that several factors can give rise to the enhancement in quantum sensing for a free fermionic model.

\emph{\textbf {Localization-delocalization sensor.} }  In this class of sensors, the Stark probe is described by the Hamiltonian
\begin{equation}
    H_{\text{FH}}^\text{spinless} = -\sum_{j} (c_{j}^{\dagger} c_{j+1} + \text{H.c.}) + \sum_{j} h_j\, c_{j}^{\dagger}c_j,
\end{equation}
where $\hat{c}^{\dagger}_j$ ($\hat{c}_j$) is the creation  (annihilation) operator at site $j$ of a spinless fermion and the onsite potential varies linearly as $h_j = h j$. This system exhibits a localization-delocalization transition as $h \to 0$  in the thermodynamic limit. Near this criticality, estimating weak values of the field strength $h$ leads to QFI scaling beyond the Heisenberg limit (HL). Specifically, in the single-particle case, it was demonstrated that the QFI scales with system size ($L$) as $\mathcal{F}_h \sim L^6$, and in the half-filled many-body case, it scales as $ \mathcal{F}_h \sim L^{4.2}$ ~\cite{He2023}. Remarkably, by introducing a nonlinear term to the on-site potential, this scaling can be significantly enhanced, enabling even greater sensitivity in parameter estimation~\cite{Yousefjani2025a}. Alternatively, instead of a linear potential, one can consider a quasiperiodic potential of the form $ h_i = h \cos(2\pi\omega j) $, where $\omega$ is the ratio of two consecutive Fibonacci numbers. This system also exhibits a localization-delocalization transition, but at a finite critical value $h = 2$ in the thermodynamic limit. The resulting model is known as the Aubry-André (AA) model. Leveraging the criticality in the AA model, the QFI saturates the Heisenberg limit (HL), scaling as $\mathcal{F}_h \sim L^2$, for both the single-particle and half-filled many-body cases~\cite{Sahoo2024a}. Recent studies have explored hybrid models that combine both AA and Stark potentials. Notably, at the critical point of the AA model, it is possible to achieve better precision in estimating the Stark field compared to the pure Stark model~\cite{Sahoo2025, Sahoo2024b}. 

{\emph{\textbf {Topological sensor.}}} Fermionic topological phase transition (TPT) turns out to have several features which are different than phase transitions with symmetry-breaking and can provide quantum benefits in sensing which are robust against local disturbance.
Specifically, a paradigmatic example exhibiting TPT is the one-dimensional Su-Schrieffer-Heeger (SSH) model, described by the Hamiltonian,  
\begin{equation}
    H_{\text{SSH}} = - \sum_{j} \left( J_1\, b^{\dagger}_j a_j + J_2\, a^{\dagger}_{j+1} b_j + \text{H.c.} \right),
\end{equation}
where $J_1$ and $J_2$ denote the coupling between the internal state on the same site and the adjacent sites, respectively. The fermionic operators $a_j$ and $b_j$ correspond to two internal states within each unit cell. 
Setting $ \lambda = J_1 / J_2 $, the system undergoes a topological phase transition (TPT) at $\lambda = 1$. For $\lambda < 1 $, the SSH chain supports topologically protected edge states.
By varying $\lambda$, the quadratic to constant scaling for estimating the parameter is reported, thereby ensuring HL near TPT. Moreover, the measurement basis, namely the precision measurement is recognised which leads to saturation of the  Cramér-Rao bound~\cite{Sarkar2022}. 

\begin{figure*}
    \centering    
    \includegraphics[width=\linewidth]{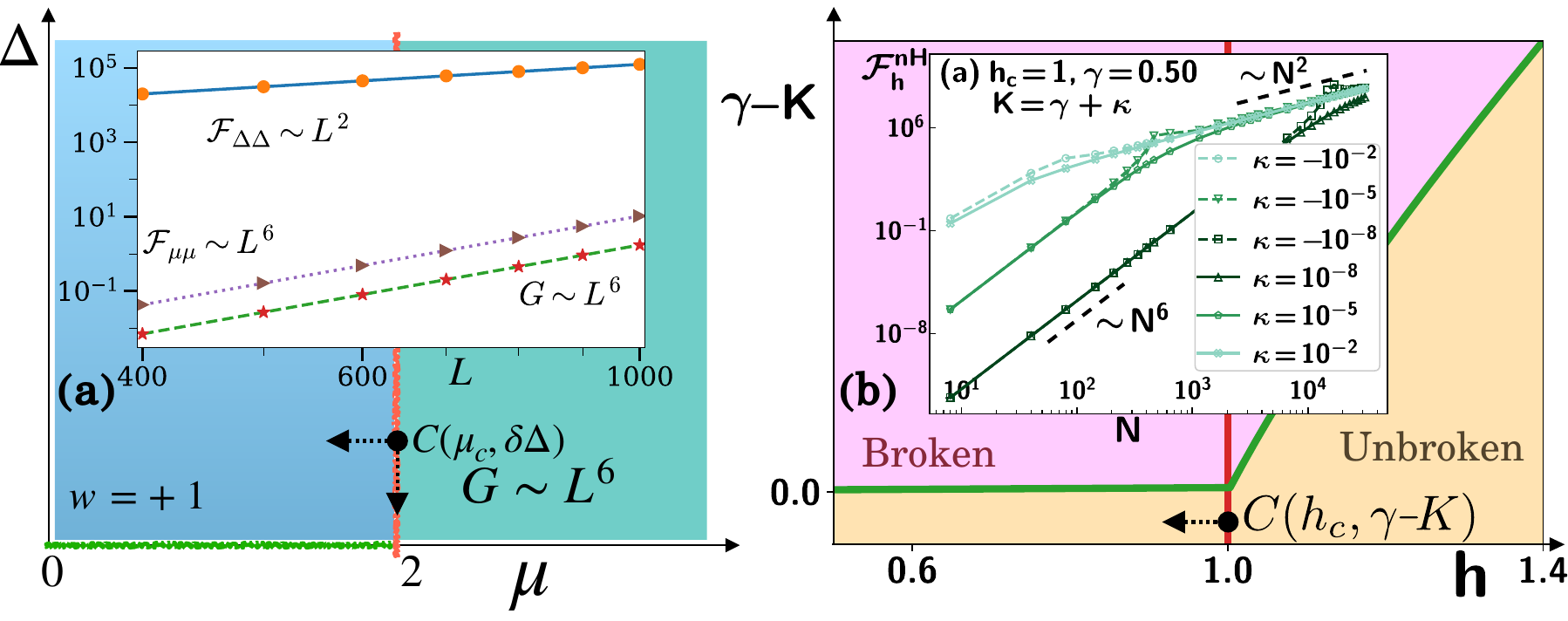}    \caption{\emph{\textbf{Criticality enhanced sensing in Fermionic and Non-hermitian spin systems.}} (a) The topological phase diagram of the one-dimensional Kitaev model (given by Eq.~\eqref{eq:kitaev}) is presented with $\mu = \mu_c = 2$ being the gapless critical line (depicted with red) separating the topologically trivial phase (depicted with green) from the topological phase (depicted with blue) with winding number $w=+1$. The multi-parameter QFI, $G= (\text{Tr}[\mathbb{F}^{-1}])^{-1}$ ($\mathbb{F}$ being the QFI-matrix for multi-parameter estimation) of a point $C$ at the critical line with $\delta\Delta\rightarrow 0$ has super-HL scaling. 
    In the inset, the system-size scaling of the QFI-s corresponding to estimating $\mu$, $\Delta$ and both simultaneously, which are given by $\mathcal{F}_{\mu\mu}$, $\mathcal{F}_{\Delta\Delta}$ and $G$ respectively for $\mu = \mu_c = 2$ and $\Delta = 10^{-7}$.  
    (b) Non-Hermitian XY model with $KSEA$ interactions $K$ and non-Hermitian anisotropy $\gamma$ have rotation time $\mathcal{RT}$-symmetry. It can be mapped to Eq.~\eqref{eq:kitaev} with $\mu=2h$ and $H_\Delta$ extended to $H_{\gamma,K}=\sum_j i(\gamma-K)c_j^\dagger c_{j+1}^\dagger+i(\gamma+K)c_j c_{j+1}$. For the gapless critical line at $h=1$ (marked by red), the broken phase is separated from the unbroken phase by exceptional points (marked by green) shown for fixed $K$. The ground states shows super-HL scaling for sensing magnetic field $h$, as $\gamma-K\to 0$, showing advantage via competition between Hermitian and non-Hermitian interactions. Figures (a) and (b) are adapted from Ref.~\cite{Mondal2024} and~\cite{agarwal2025} respectively.}
    \label{fig:kitaev}
\end{figure*}

The Kitaev chain is a topological model that does not conserve particle number due to the presence of a pairing term. The Hamiltonian of the system is given by $ H_\text{Kitaev} = H_s + H_{\Delta} $, where
\begin{align}
    \label{eq:kitaev}
    H_s &= -\sum_{j = 1}^{L} \left(c_j^\dagger c_{j+1} + \text{H.c.} \right) 
    - \mu \sum_{j = 1}^L \left( c_j^\dagger c_{j} - \frac{1}{2} \right), \nonumber \\
    H_{\Delta} &= \frac{\Delta}{2} \sum_{j = 1}^{L} \left( c_j^\dagger c_{j+1}^\dagger + \text{H.c.} \right),
\end{align}

where $\mu$ and $\Delta$ denote the chemical potential (which controls the number of particles), and the strength of the $p$-wave superconducting pairing between neighboring sites respectively. This system exhibits three topologically distinct phases separated by the critical lines $\mu = \pm 2$ and $\Delta = 0$: a trivial phase with $w = 0$ for $|\mu| > 2$, a topologically non-trivial phase with $w = 1$ for $|\mu| < 2$ and $\Delta > 0$, and another non-trivial phase with $w = -1$ for $|\mu| < 2$ and $\Delta < 0$. Ref.~\cite{Mondal2024} demonstrates that HL precision in measuring the superconducting coupling can be achieved by preparing the system at or near the multicritical point, even when the parameter variations are restricted to the critical lines, i.e., without necessitating a gap-to-gapless transition(see Fig.~\ref{fig:kitaev}(a)). 
Moreover, instead of nearest-neighbour interactions, the long-range Kitaev chain can also provide the super HL, as demonstrated in Ref.~\cite{Yang2022}. Further, the topological Rice-Mele model 
of spinless fermions is identified as a potential candidate for building optimal local thermometer~\cite{Srivastava2025},
which are capable of operating in the sub-nanokelvin temperature regime.

{\emph{\textbf {Non-Hermitian lattice sensor.}}} 
In certain physical scenarios, such as particle gain or loss in open quantum systems, non-Hermitian Hamiltonians naturally arise, unlike Hermitian operators, they typically have complex eigenvalues, and their right and left eigenvectors are distinct, and they form a biorthogonal set. Interestingly, non-Hermitian Hamiltonians can still possess entirely real spectra if they exhibit parity-time ($\mathcal{PT}$) symmetry, i.e., when the Hamiltonian $H$ is invariant under the combined operations of parity ($\mathcal{P}$) and time-reversal ($\mathcal{T}$), i.e., $[H, \mathcal{PT}] = 0$. $\mathcal{PT}$-symmetric non-Hermitian systems turns out to provide enhanced sensitivity near the exceptional points (EPs)\footnote{At the exceptional point, the system undergoes a transition from a broken (where all the eigenvalues are complex) to an unbroken (in which energy spectrum becomes real) regime. Moreover, at the exceptional points, two or more eigenvalues and their corresponding eigenvectors coalesce. }~\cite{Liu2016b}. Further, the energy spectrum both in the broken and unbroken regimes of the quantum $\mathcal{PT}$-symmetric Hamiltonian is observed by using weak measurement and in the broken region, the enhanced sensitivity is studied in experiments ~\cite{Yu2020, Yu2024b}. 
One of the first proposal in this direction demonstrates that  a microcavity sensor can provide higher precision for single-particle detection at the EPs~\cite{Wiersig2014}.  Further works include an EP-based amplifier operating near the lasing threshold with heterodyne signal detection~\cite{Zhang2019}, enhanced sensitivity in optical microcavities using nanoscale scatterers~\cite{Chen2019}, higher-order EPs~\cite{Hodaei2017}, and increased sensitivity through laser noise in gyroscopes~\cite{Wang2020b}. $\mathcal{PT}$-symmetric electromechanical accelerometers have also shown to have EP-enhanced performance~\cite{Kononchuk2022}(see also Refs.~\cite{Chen2017, Wiersig2020, Langbein2018}). Notably, non-Hermitian sensing schemes that do not rely on EPs have also been proposed, see Ref.~\cite{Xiao2024}.  

A well-known non-Hermitian model without disorder is the Hatano-Nelson (HN) Hamiltonian, given by
\begin{equation}
\label{HN}
H_{\text{HN}} = \sum_{j} \left( t_R\, c^\dagger_{j+1} c_j + t_L\, c^\dagger_{j} c_{j+1} \right),
\end{equation}
where $t_R \neq t_L$. Under open boundary conditions, the left and right eigenstates of this Hamiltonian tend to localize exponentially at the system’s boundaries and the eigenvalue spectrum becomes exponentially sensitive to boundary conditions~\cite{Edvardsson2022}, known as the non-Hermitian skin effect \cite{Gohsrich2025}. 
Interestingly, exploiting this skin effect QFI is shown to get increased~\cite{Sarkar2024, Arandes2025}. 
Additionally, in the presence of transverse magnetic-field whose strength is to be estimated, $\mathcal{RT}$-symmetric\footnote{$\mathcal{RT}$-symmetry represents combined rotation ($\mathcal{R}$) and time ($\mathcal{T}$) symmetry of the Hamiltonian, i.e., $[H,\mathcal{RT}] = 0$. } non-Hermitian XY chain with Kaplan-Shekhtman-Entin-Aharony ($KSEA$) interaction, have been shown to surpass the HL~\cite{agarwal2025} for finite system-sizes, near the tri-critical point where exceptional and critical lines overlap (see Fig.~\ref{fig:kitaev}(b)).

Recent works~\cite{McDonald2020, Lau2018, Guo2021,Sarkar2025b} have demonstrated quantum-enhanced sensing in non-Hermitian system. Beyond specific models,  Ref.~\cite{Ding2023} establishes the fundamental sensitivity limits for non-Hermitian quantum sensors, while Ref.~\cite{Sarkar2024} finds that the enhanced sensitivity is related to the closing of the point gap, a uniquely non-Hermitian energy gap with no Hermitian analogue.

\subsection{The Bose-Hubbard model}

Let us now move to the Bose-Hubbard (BH) model, described by the Hamiltonian,
\begin{equation}
H_{\text{BH}} = -t \sum_{\langle i,j \rangle} ( b_i^\dagger b_j + \text{H.c.} ) + \frac{U}{2} \sum_i n_i (n_i - 1) - \mu \sum_i n_i,
\label{eq:BH}
\end{equation}
where $b_i$ ($b_i^{\dagger}$) is the bosonic annihilation (creation) operator, and $n_i$ is the number operator at the $i$-th site and $t$, $U$ and $\mu$ represent the hopping, on-site interaction and chemical potential respectively.
The Hamiltonian possesses a global $\text{U}(1)$ symmetry, i.e., it remains invariant under the transformation $b_i \to e^{i\theta}b_i$, where $\theta$ is an arbitrary global phase. At zero temperature ($T = 0$), the system undergoes a quantum phase transition between two distinct phases by varying $U/t$ -- the superfluid ($t\gg U$) and the Mott insulating phase. The former one is characterized by long-range phase coherence, spontaneous breaking of $\text{U}(1)$ symmetry, nonvanishing compressibility, and a gapless excitation spectrum. Conversely, in the Mott insulating phase ($t \ll U$), the interaction strength dominates, leading to localization of bosons and preservation of $U(1)$ symmetry. 
A landmark experimental observation of this quantum phase transition was reported for ultracold $^{87}\mathrm{Rb}$ atoms confined in a three-dimensional optical lattice ~\cite{Greiner2002}. 

{\emph{\textbf{Interferometric quantum sensing in BH.} } } In an interferometric quantum sensing setup, the objective is to estimate a phase that is encoded into a quantum state via a unitary operation. Quantum-enhanced sensitivity in this interferometric framework typically relies on using specially prepared initial states such as GHZ-type entangled states, NOON states, and squeezed states. However, generating these highly entangled states experimentally poses significant challenges and requires substantial resources. A theoretical protocol for dynamically generating and storing a robust, highly entangled GHZ state in a system of bosonic atoms confined in a one-dimensional optical lattice has been proposed, along with a method for detecting entanglement in such systems \cite{Maciej2023}. Moreover, numerous theoretical studies have demonstrated the feasibility of generating GHZ-like states \cite{booker2020, krutitsky2017, Eichler2014, Sharma2011, wendt2011} and NOON states \cite{Wittmann2023, Yannouleas2019, Compagno2017, Watanabe2010, Watanabe2012} within the framework of the Hubbard model. \\

One important feature of the BH model is that particles can form bound states when they gather at the same site due to strong onsite interactions ($U$). Ref \cite{Compagno2017} shows that local impurity can help to generate NOON state between the edges of finite lattice. In particular, this method of creating NOON state can be used to estimate external local field in Mach-Zehnder interferometer. In optical lattice experiments, this setup can be created using single-atom control techniques, starting from a Mott-insulator phase \cite{Weitenberg2011, Valiente2010}. Further, the precision of gyroscope, useful for modern navigation system, can also be enhanced with the help of entangled NOON state \cite{Cooper2010}. On the other hand, by considering an atomic gyroscope consisting of $N$ ultracold atoms of mass $m$, confined in a three-site optical lattice arranged in a ring of circumference $L$, optimal initial state identified that maximizes and thus the highest phase estimation precision is shown to achieve using either a squeezed entangled state or an entangled even squeezed state as the input ~\cite{shao2020}.

\begin{figure}
    \centering    
    \includegraphics[width=\linewidth]{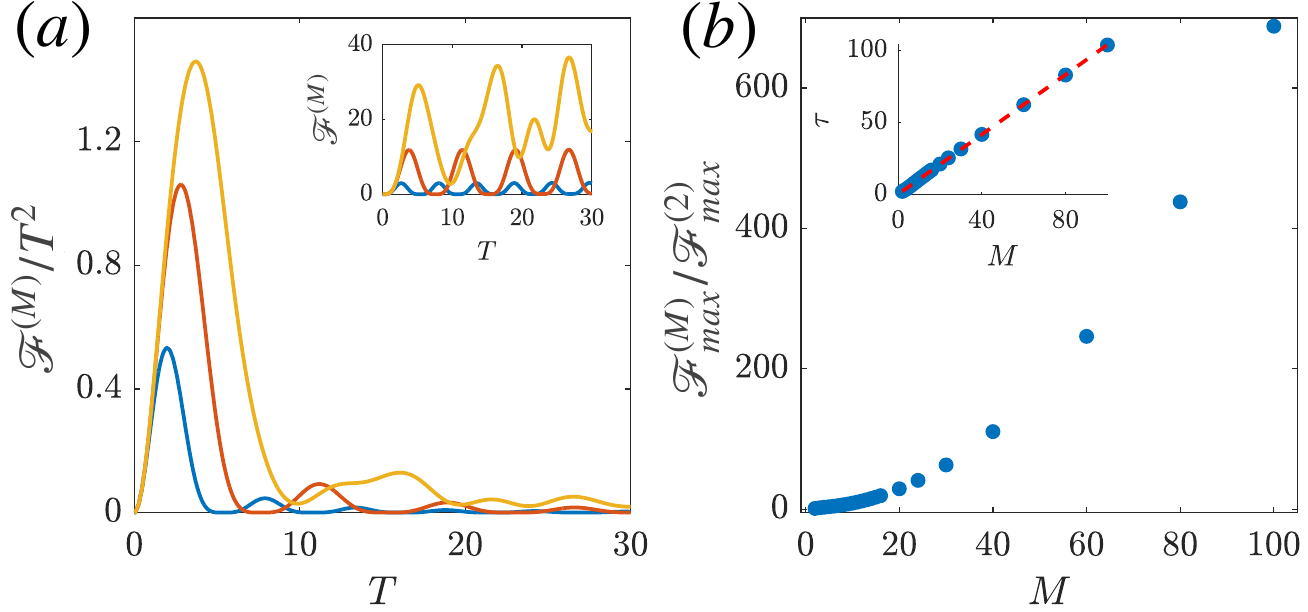}    \caption{\emph{\textbf{Parameter estimation for an initial Fock state under a periodic drive.}} 
    {\color{black}{(a) The time evolution of the QFI, $\mathcal{F}^{(M)}$, for a Fock state as the initial state is shown here. The results are presented for total lattice sites $M = 2$ (blue), $3$ (red), and $4$ (orange). Here, $\mathcal{F}^{(M)}$ is scaled by $T^2$, and the peak value is defined as $\mathcal{F}^{(M)}_{\text{max}} = \max\left\{\frac{\mathcal{F}^{(M)}}{T^2}\right\}.$ The inset displays the unscaled QFI, which exhibits an oscillatory behavior as a function of time $T$. (b) The growth of the normalized peak QFI, $\mathcal{F}^{(M)}_{\text{max}} / \mathcal{F}^{(2)}_{\text{max}}$, shows a quadratic dependence on the system size $M$ in the limit $M \gg 1$. The inset illustrates that the $\tau$, the time at which $\mathcal{F}^{(M)}_{\text{max}}$ reaches its maximum value, increases linearly with $M$. The figure is taken from Ref.~\cite{Pelayo2023}.}} }
    \label{fig:distributed_qs}
\end{figure}

Recent studies suggest that instead of using a constant onsite potential in the last term of BH model in Eq.~\eqref{eq:BH}, one can consider an additional linear potential $\mu \sum_i i n_i$, where $\mu$ is the strength of the tilt and this system is known as the tilted Bose-Hubbard (TBH) model \cite{Debnath2025, Pelayo2023}, realizable using ultracold atoms trapped in a 1D optical lattice \cite{Poli2011, Miyake2013}. To estimate the parameter $\mu$ in the TBH system, one approach to reach the HL is by preparing a NOON state, although they are not robust against losses and hence preparation of high fidelity NOON state in the laboratory is difficult. To overcome this situation, in the system Ref.~\cite{Pelayo2023} demonstrates that starting from a Fock state the periodic drive which depends on the parameter to be estimated can provide high QFI, ensuring improved high sensitivity (see Fig.~\ref{fig:distributed_qs}).

{\emph{\textbf{Spin squeezing in BH.}}}   In Sec .~\ref {metro_BEC}, we have a detailed discussion on the spin-squeezing technique, which allows for very precise measurements that can go beyond the shot noise limit (SNL). {\color{black}{There are mainly two theoretical models. The first one is the OAT case, where the spin squeezing parameter scales as $\xi \propto N^{-1/3}$ at time $\chi t \propto N^{-2/3}$, with $N$ being the total number of quantum particles in the system while the second one is the TACT model, which allows one to reach the HL, where $\xi \propto N^{-1/2}$ at $\chi t \propto N^{-1} \log(2N)$. }}

By considering mapping of the BH model to an effective single-particle system, spin squeezed state was created by adopting a fast-adiabatic-like preparation method~\cite{Juli2012} while the optimal control theory approach was taken to produce cat-like states~\cite{Lapert2012}.  The OAT dynamics can be simulated using a two-component BH model through contact interactions among bosons in the superfluid phase~\cite{Plodzien2020}. In contrast, the TACT dynamics can be realized by incorporating contact as well as dipolar interactions in the two-component BH model~\cite{Dziurawiec2023}. The corresponding Hamiltonian for the two-component BH model with dipolar interactions, where $N$ bosons occupy two internal states $|\uparrow\rangle$ and $|\downarrow\rangle$, is given by $H = H_{\text{BH}} + H_d$. Here, $H_{\text{BH}}$ represents the two-component BH Hamiltonian, and the dipolar interaction term $H_d$ is given by
$H_d = \sum_{j,k \neq j} W_{jk} \left( S^z_j S^z_k - 2 S^x_j S^x_k + S^y_j S^y_k \right),$
where $S_j^{x,y,z}$ are the spin operators in the $x$, $y$, and $z$ directions, $W_{jk} = \frac{W_0}{|j - k|^3}$, with $W_0 = \frac{\gamma^2 \hbar^2 \mu_0}{4\pi d^3},$ being a dipole-dipole coupling constant, which depends on the gyromagnetic ratio $\gamma$, the reduced Planck constant $\hbar$, the vacuum permeability $\mu_0$, and the lattice constant $d$. Ref.~\cite{Dziurawiec2023} demonstrates that the above system in the superfluid phase can simulate the TACT dynamics. By varying anisotropy parameter, both OAT and TACT dynamics can be simulated in this model and for a weak anisotropy, it is possible to achieve the HL of squeezing \cite{Bonkhoff2023, Yuste2013, Juli2012, Roscilde2010, en2023, Adhikary_2018, Quijandría_2015, ho2009squeezing}.

{\emph{\textbf{Real-time feedback control. } } } As discussed in previous sections, critical quantum sensing relies on preparing a quantum state near a critical point to achieve quantum-enhanced sensitivity, although it faces significant challenges during realization. To overcome this issue, a two-step adaptive scheme is proposed which can surpass the shot noise limit \cite{Salvia2023}. They focus on the estimation of an unknown parameter $\lambda$, which is assumed to follow a prior distribution $p_0(\lambda)$. First, $\epsilon m$ ($\epsilon \ll 1$) number of identical measurements are performed for some direction chosen according to prior knowledge and then $\lambda$ can be estimated. In the second step, the remaining copies are measured for a configuration which is chosen accordingly to the assumption that the system is at criticality. The method involves performing $m$ consecutive measurements on the ground state $\rho(\lambda, \vec{s})$, where the Hamiltonian depends on an externally parameter $\vec{s}$ which can be adjusted during the measurement process to enhance sensitivity. After the $k$-th measurement, the posterior distribution is updated using Bayes' rule, based on the conditional probability of the observed outcome. Following each measurement, an estimator is constructed to assign a value to the unknown parameter $\lambda$ based on the accumulated data. The performance of the estimation process is evaluated using the expected mean square distance (EMSD) as a figure of merit. Within this framework, the classical Fisher information (CFI) can also be computed, as it depends on the conditional probability distribution of the measurement outcomes. By optimizing over all possible measurements, the maximum achievable CFI saturates to the QFI. The effectiveness of the method is applied on the 2D BH model to estimate the hopping parameter $t$, using the on-site interaction $U$ as a control parameter and $\mu = 1/2$. This model can be implemented in Josephson junction arrays, where $U$ can be tuned by adjusting the junction capacitance~\cite{Bradley1984}. At the critical point $t_c \simeq 0.06U$, the QFI is shown to scale with the system size $L$ as $L^{1.34}$. Note further that several studies have demonstrated the advantages of enhanced sensing through real-time feedback control in various platforms \cite{nugroho2019control, Berni2015, Sootla2016, Brakhane2012}.

\begin{figure}[t!]
    \centering
    \includegraphics[scale = 0.35]{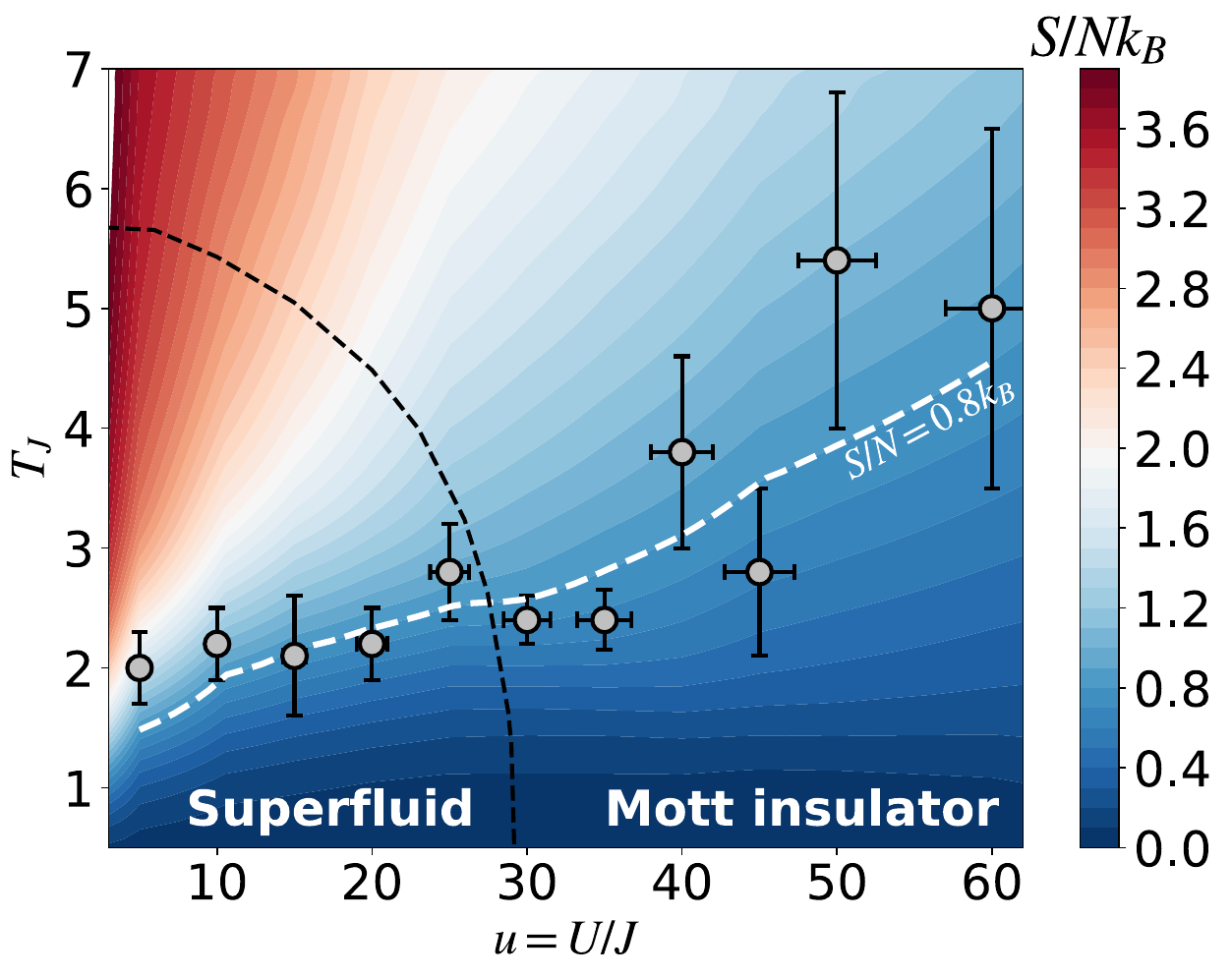}
    \caption{\emph{\textbf{Thermometry in BH model.}} The false-color plot shows the variation of entropy per particle, $S/Nk_{B}$, as a function of the interaction parameter $u = U/J$ and the dimensionless temperature $T_J = k_B T / J$ for the 3D Bose-Hubbard model. The open circles represent values of $T_J$ extracted using momentum-space density, $\rho(k)$, thermometry. The white dotted line indicates an isentropic trajectory corresponding to $S/N = 0.8k_B$, while the black dotted line marks the critical temperature at unit filling. This figure is taken from Ref.~\cite{Carcy2021}.
 }
    \label{fig:Bose_hubbard}
\end{figure}

\emph{\textbf{Temperature sensing in BH.}}  
The goal is to estimate temperature for the trapped 3D BH model which cannot be predicted by using density in the momentum space. It is done by computing the Fisher information obtained in several experimental attempts. Interestingly, Fisher information, and hence the precision in estimation get maximized near the phase transition which is demonstrated by using $^4\text{He}$ atoms. Further by using quantum Monte-Carlo simulation, energy and specific heat of the model can be obtained which leads to the behavior of entropy (see Fig.~\ref{fig:Bose_hubbard}) when particle number is fixed, i.e., for a canonical ensemble~\cite{Carcy2021}.   

In addition, several studies have proposed different methods for measuring temperature in the BH model, each offering specific advantages and demonstrating experimental reliability \cite{Suthar2022, Xu2019, Pohl2022, Adebanjo_2022, Masaki-Kato2022, Emanuel2023, PhysRevB.102.214503, PhysRevA.104.043320}.\\

\subsection{Sensing in spin systems }

Let us move to the estimation of parameter involved in quantum spin models. The first part concentrates on the scaling of QFI with system-size when the system is either at zero temperature or finite temperature while in the second part, the parameter is estimated during the dynamics of the system.

 \begin{figure*}
    \centering    
    \includegraphics[width=\linewidth]{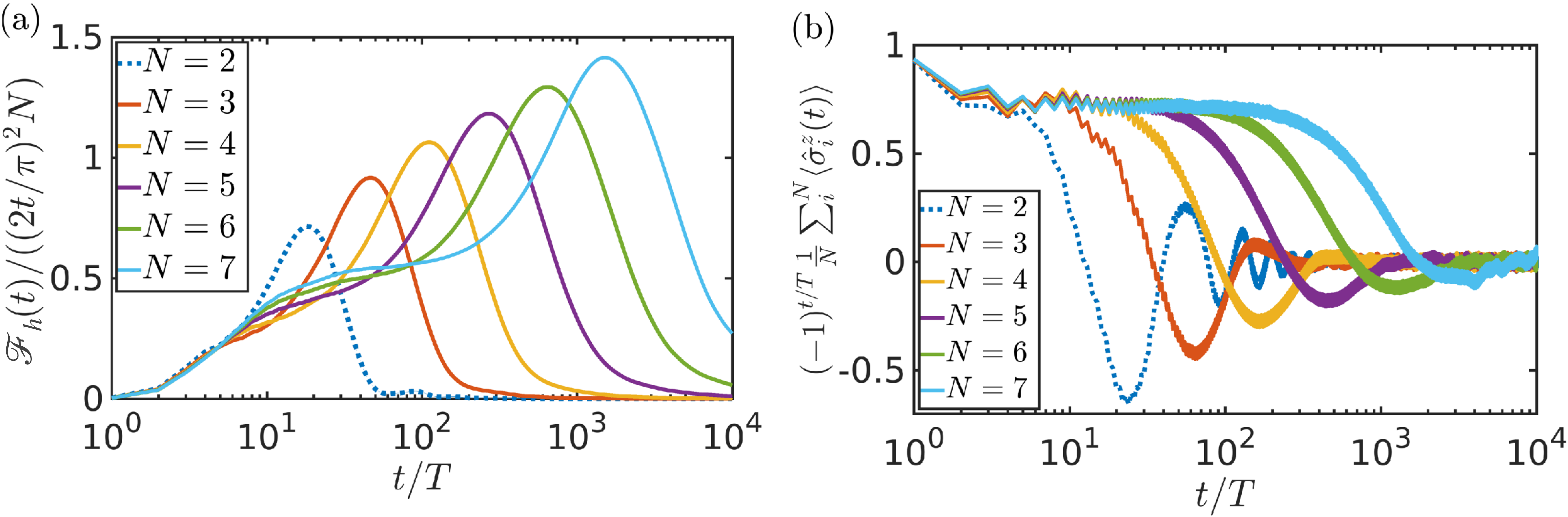} 
    \caption{\emph{\textbf{Ac field sensing by floquet time crystal}} 
    {\color{black}{(a) The QFI dynamics as a function of the FTC sensor size $N$ at resonance $\omega_h = \pi / T$ are presented here. In the linear-response regime, the time evolution of the QFI exhibits an initial transient growth, after which it reaches a plateau corresponding to the SQL scaling in time ($\sim t^2$). At later times, the QFI surpasses the SQL, continuing to grow faster than quadratically until the system eventually thermalizes at times that increase exponentially with $N$. In this intermediate regime, the maximum QFI displays super-quadratic scaling in time,  $\frac{\mathcal{F}_h(t)}{N} \sim t^{2\beta(t)}, \quad \text{with } \beta(t) > 1.$ (b) The corresponding magnetization dynamics are shown. Following an initial transient period, the magnetization reaches a plateau whose duration coincides with that of the QFI plateau. At longer times, dephasing induced by many-body localization (MBL) causes the magnetization to gradually decay toward its thermal value. The thermalization time---which grows exponentially with $N$---matches the time window during which the QFI exceeds the SQL scaling. The analysis is performed for the parameter values $J = b_z = 0.25, \quad b_x = 0.025, \quad \theta = 0, \quad \phi = 2.8, \quad T = 1, \quad 
\omega_h = \frac{\pi}{T}, \quad h \to 0.$  The figure is taken from Ref.~\cite{Iemini2024}. }}  }
    \label{fig:floquet_qs}
\end{figure*}

\subsubsection{Equilibrium quantum sensing}  
As discussed earlier in Sec.~\ref {critical_sensing_in_lattice}, quantum criticality can enhance the precision of parameter estimation beyond classical limits. 
A simplest spin model that undergoes quantum phase transition, is the nearest-neighbor transverse Ising model~\cite{Sachdev2011}, given by, 
\begin{equation}
\label{Ising_chain}
H_\text{TIsing} = -J\sum_{i} \sigma_i^{x} \sigma_{i+1}^{x} + h \sum_{i} \sigma^z_{i},
\end{equation}
where $J$ denotes the nearest-neighbor coupling strength, $h$ is the transverse magnetic field, and $\sigma_i^{x,z}$ are the Pauli spin operators at site $i$. 
Near the critical point $\frac{h}{J}\to \pm 1$, the QFI for estimating the parameter \(h\) scales quadratically with the system size $L$, i.e., $\text{QFI} \sim L^2$, thus achieving the HL~\cite{Zanardi2008a, Skotiniotis2015}. 
However, as discussed before, this quantum enhancement is exciting although hard to realize and hence adaptive measurement scheme 
involving real-time feedback control 
can be used to attain 
quantum advantage beyond the SQL~\cite{Salvia2023}.

Beyond the transverse Ising model, plethora of quantum spin chains have been explored
to estimate magnetic field, see Ref.~\cite{Montenegro2021} for the Ising chain, for XY spin chains Ref.~\cite{liu2013quantum,Mihailescu2025a}, systems with Dzyaloshinskii–Moriya interactions~\cite{Ozaydin2015a, Adani2024, BenHammou2024}, the Heisenberg XY model~\cite{Bakmou2019}, and the antiferromagnetic Heisenberg chain with uniaxial anisotropy~\cite{Lambert2019}. Studies have also extended to multiparameter estimation near criticality~\cite{Fresco2022, jiang2021}, and fermionic systems such as the $t\text{-}U\text{-}J$ model~\cite{Lucchesi2019}. Recent works have demonstrated that criticality-driven metrology schemes can even be applied to phenomena like the quantum Hall effect~\cite{Giraud2024}, and higher-dimensional spin chains~\cite{Singh2024}.

The role of long-range interactions in quantum sensing is another important direction of research. 
Ref.~\cite{Lorenzo2018} shows that in the presence of classical long-range correlations, one can also achieve HL. Other studies include sensing in fully connected systems~\cite{Garbe2022}, and models such as the long-range Kitaev chain~\cite{Yang2022}, XY chain~\cite {Mihailescu2025b}, and Stark spin chains~\cite{Yousefjani2023} that can surpass SQL. In a protocol~\cite{Yoshinaga2021}, it was also found that although quasi long-range interacting Hamiltonian can surpass SQL although true long-range interacting system fails to provide quantum advantage~\cite{Mothsara2025}. 
A recent review \cite{reviewMontenegro2024} discusses quantum sensing in spin chains with details.

\subsubsection{Dynamical quantum sensing}
Here, we broadly discuss three types of dynamical sensing protocols: Floquet systems, time crystals, and quantum many-body scars that help to achieve quantum benefits. 

\emph{\textbf{Floquet systems.}} In many-body quantum spin systems, it has been observed that Floquet-driven systems can enhance quantum sensing~\cite{Mishra2021, Mishra2022}. In the XY model, $\hat H_\text{fl}(t) = -\frac{J}{2} \sum_{i = 1}^L\left[ \left(\frac{1+\gamma}{2}\right)\sigma^x_i\sigma_{i+1}^x + \left(\frac{1-\gamma}{2} \right)\sigma^y_i\sigma_{i+1}^y \right] -\frac{[h_0 + h(t)]}{2}\sum_{i=1}^N \sigma_i^z$, the external field is driven by the time-dependent field $h(t) = h_1\sin(\omega t)$ with periodic boundary conditions and the driving time period being $\tau = 2\pi/\omega$. 
At the critical point $h_0 = J$, without time-evolution, QFI $ \sim L^2$, while away from criticality, the QFI $ \sim L$. 
In the situation where part of the system is accessible, i.e., among the total $L$ spins, only $N$ spins can be used to estimate the parameter. The QFI for such partially accessible equilibrium sensing falls below the HL. Ref.~\cite{Mishra2021} exhibits that the Floquet driving of the form  $h(t)$ can restore the HL and even go beyond HL. According to the Floquet formalism, for a periodic Hamiltonian $\hat H_\text{fl}(t+\tau) = \hat H_\text{fl}(t)$, the states $|\psi(n\tau)\rangle$ can be obtained using $|\psi(n\tau)\rangle = \sum_i e^{-e_in\tau} |e_i\rangle\langle e_i|\psi(0)\rangle$, with $|\psi(0)\rangle$ being the initial state, $\{|e_i\rangle\}$ being the eigenstates with $e_i$ eigenvalues of the unitary operator for the evolution of the first period $U(\tau) = \mathcal Te^{-i\int_0^\tau \hat H_\text{fl}(t) dt}$. Here $\mathcal T$ is the time-order operator. For sensing, the steady state of the partially accessible system is used as a probe. 


\emph{\textbf{Discrete time crystals.}} Time crystals is a phase of matter where the system exhibits time translation symmetry-breaking~\cite{Sacha2018, Zaletel2023}. Time crystal phase has been shown to be useful for enhanced quantum sensing~\cite{Lyu2020, Biswas2025a, Biswas2025b, Yousefjani2025b}. In Ref.~\cite{Iemini2024}, a spin system is considered with periodic kicks at intervals of $T$, given by $\hat  H_\text{tc} = \sum_i\left[J_i \sigma_i^z \sigma_{i+1}^z + \sum_{\alpha = x,z}b^\alpha_i\sigma^\alpha_i -\frac{\phi}{2}\sum_{n = -\infty}^{+\infty} \delta(t-nT)\sigma^x_i\right]$. The system evolves under disordered interactions such that $J_i$, $b_i^z$ and $b_i^x$ are chosen from uniform distribution. The system without the periodic kicks, attains a many-body localized (MBL) phase, while with the periodic kicks, it stabilizes in the floquet time crystal (FTC) phase. This is used to sense the magnitude $h$ of small ac fields given by $\hat V_\text{tc} = h\sin(\omega_ht + \theta)\sum_i \sigma_i^z/2$. For sensing, the system evolves under the Hamiltonian $\hat H_\text{tc}+\hat V_\text{tc}$ with $T\sim \pi/\omega_h$. The QFI $\mathcal{F}(t)$ shows SQL growth in time initially, while later on, it attains higher than SQL scaling in time $\mathcal{F}(t)/N \sim t^{2\beta(t)}$ with $\beta(t)>1$. The collective interactions of the system stabilize its dynamics and make it robust against noise and imperfection (see Fig.~\ref{fig:floquet_qs}).     

\begin{figure*}
    \includegraphics[width=0.95\textwidth]{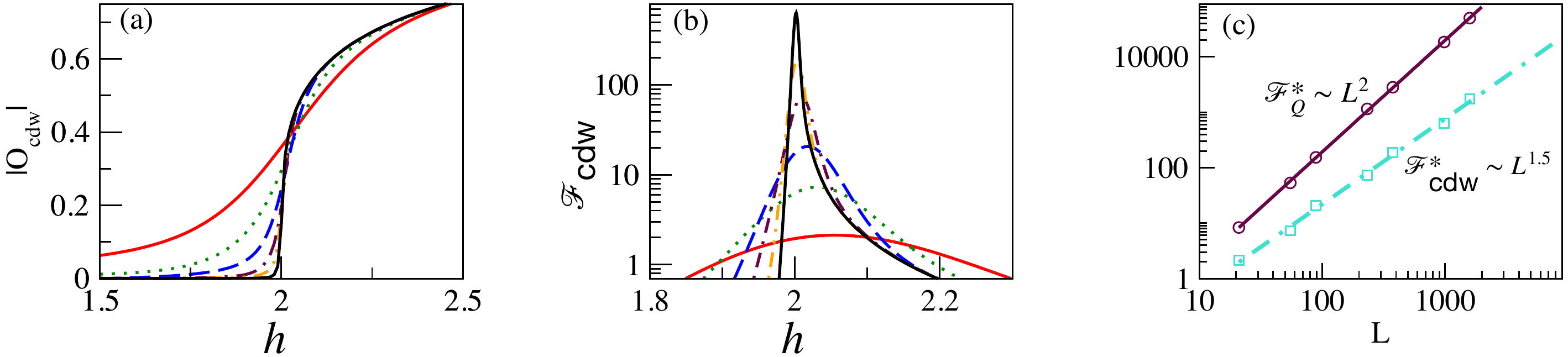}
\caption{{\bf OFI in quasi-periodic lattice for single particle case:} 
{\color{black}{(a) The modulus of charge density wave (CDW) order parameter, $|O_{\mathrm{CDW}}|$, as a function of the quasiperiodic potential strength $h$.  (b) The OFI, $\mathcal{F}_{\mathrm{CDW}}$, corresponding to the operator $\hat{O}_{\mathrm{CDW}}$, is plotted against $h$ for the same set of system sizes as in panel (a).  Near the transition point $h \sim 2$, $\mathcal{F}_{\mathrm{CDW}}$ increases with the system size $L$, indicating enhanced sensitivity of the system to changes in $h$.  (c) The circles and squares represent the maximum QFI, $\mathcal{F}_Q^{*}$, and the maximum OFI, $\mathcal{F}_{\mathrm{CDW}}^{*}$, respectively, as functions of $L$. 
The solid line shows a best fit of $\mathcal{F}_Q^{*} \sim L^2$, while the dashed line indicates the best fit of $\mathcal{F}_{\mathrm{CDW}}^{*} \sim L^{1.5}$.  This scaling demonstrates that the OFI associated with the CDW operator exceeds the SQL, revealing a quantum-enhanced precision in estimating the parameter $h$. 
The system sizes $L = 21$ (solid red), $55$ (dotted), $89$ (dashed), $233$ (dash-dotted), $377$ (dash-double-dotted), and $987$ (solid black) are used consistently across panels (a) and (b). The figure is taken from Ref.~\cite{Sahoo2024a}.  }}}
    \label{fig:OFI}
\end{figure*}

\emph{\textbf{Quantum scars.}} A quantum many-body system is said to thermalize if the reduced density matrix of any small subsystem goes to the Gibbs thermal state, known as the eigenstate thermalization hypothesis (ETH)~\cite{Deutsch2018}. In certain many-body systems, there exist certain states that do not follow ETH, these states are known as quantum many-body scars (QMBS)~\cite{Serbyn2021,Moudgalya2022}. Recently, it has been demonstrated that QMBS can provide an advantage in quantum-enhanced sensing~\cite{Dooley2021, Desaules2022, Dooley2023,Guo2023b}. Ref.~\cite{Desaules2022} analytically demonstrates that QMBS with finite energy density possess HL scaling, i.e., QFI~$\sim L^2$. Furthermore, they consider the PXP model~\cite{Fendley2004,Lesanovsky2012}, and show that the MBQS have the largest QFI among all eigenstates. The scarred state is showed to have HL scaling, but at large times,  the QFI starts to drop. Thus, despite the non-exact nature of QMBS, the PXP model shows robust signature of super-extensive QFI scaling.

In addition to closed systems, recent research in open quantum systems has also demonstrated quantum advantage \cite{langfitt2023, Sarkar2025b,mattes2025,Wang2020,puig2024}.

{
\begin{table*}[t]
\centering
\renewcommand{\arraystretch}{1.3}

\newcolumntype{P}[1]{>{\RaggedRight\arraybackslash}p{#1}}
\newcolumntype{q}[1]{>{\centering\arraybackslash}p{#1}}

\begin{tabular}{|P{4.0cm}|P{9.4cm}|P{4.0cm}|}
\hline
\textbf{Platform / System} & 
\textbf{Parameter and Strategy} &
\textbf{Scaling / Sensitivity}\\ \hline

\multirow{2}{=}{Spin ensemble} 
& Thermometry~\cite{Salvatori2014, Ostermann2024} via critical sensing ~\cite{Frerot2018, Pavlov2023} and interferometry-based sensing using ground states \cite{Ma2009b}& HL \\
 \hline

\multirow{4}{=}{Cavity-QED/Light-matter hybrid systems} 
  &  Coupling interaction~\cite{Bina2016, Saleem2024}, magnetic fields~\cite{Ivanov2013} and Interferometry-based sensing using steady states~\cite{Paulisch2019}, & HL \\ 
  & Interferometry-based sensing using dynamical states~\cite{Pavlov2025},
  & Beyond HL \\
  \hline

\multirow{4}{=}{BECs in trap} 
& Gravimetry~\cite{Debs2011,Zhou2012, Mutinga2013, Carraz2014, Aguilera2014, Condon2019, Trimeche2019,  Szigeti2020, Ufrecht2021, Cassens2025}, inertial-sensing~\cite{Kapale2005, Pandey2019, Stolzenberg2025}, magnetometry~\cite{Gunther2005, Wildermuth2005, Wildermuth2006, Vengalattore2007, Aigner2008, Ng2013, Steinke2013, Eto2013,Muessel2014, Jasperse2017, Yang2017, Apellaniz2018,  Czajkowski2019, Yang2020, PalaciosAlvarez2022, Shukla2024}, time/frequency-sensing~\cite{Martin2013, Andre2004, Borregaard2013,Kruse2016} via atomic interferometry using squeezed states  & Beyond SQL\\
& Field strength via critical sensing~\cite{Mirkhalaf2020,Debnath2025} & Beyond SQL~\cite{Mirkhalaf2020}, HL~\cite{Debnath2025}\\
& Thermometry via imputiy based technique~\cite{Sabin2014, Olf2015, Mehboudi2019a}, time of flight measurements~\cite{Leanhardt2003, Gati2006a, Gati2006b}, critical sensing in spinor BEC~\cite{Aybar2022}  & sub nK temperature, relative error$\sim10 \%$\cite{Mehboudi2019a}, HL~\cite{Aybar2022}  \\

 \hline

\multirow{4}{=}{Non-ensemble systems (Quantum many-body systems such as Bose- and Fermi- Hubbard models, spin systems, BECs in optical lattices)} 
&  AC fields, DC fields, potentials, coupling strength and frequency via quantum criticality, such as first-order \cite{sarkar2025exponentially}, second-order \cite{Zanardi2008a, Skotiniotis2015,Montenegro2021,Mihailescu2025a, Fresco2022, Adani2024, BenHammou2024,Fresco2022, jiang2021,Lorenzo2018}, topological \cite{Sarkar2022,Mondal2024,Yang2022} and localization-delocalization \cite{Sahoo2024a,He2023,Yousefjani2025a,Sahoo2024b, Sahoo2025,Debnath2025} transitions, quantum scar \cite{Dooley2021}, spin-squeezing \cite{He2019,Hern2022,Dziurawiec2023,Dziurawiec2023,Bonkhoff2023, Yuste2013, Juli2012, Roscilde2010, en2023, Adhikary_2018, Quijandría_2015, ho2009squeezing}, Floquet driving \cite{Mishra2021, Mishra2022, Pelayo2023}, time-crystal \cite{Iemini2024, Lyu2020,Yousefjani2025b}; thermometry by density fluctuations\cite{Hartke2020, Pasqualetti2024, Carcy2021, Suthar2022}, topology \cite{Srivastava2025} & Beyond SQL \cite{Ozaydin2015a, Mishra2022}, HL\cite{Sahoo2024a,Zanardi2008a, Skotiniotis2015,Montenegro2021} \phantom{.}\cite{Mihailescu2025a,Fresco2022,Sarkar2022,Mondal2024,Yang2022} \phantom{.}\cite{Lorenzo2018,He2019,Dziurawiec2023,Bonkhoff2023} possible to go beyond HL \cite{He2023,Yousefjani2025a,Sahoo2024b, Sahoo2025, Debnath2025} \phantom{.}\cite{Mishra2021,Lyu2020,Yousefjani2025b} exponentially-enhanced \cite{sarkar2025exponentially} \\
\hline

\end{tabular}
\caption{\color{black}Summary of quantum metrology platforms, strategies, scaling behavior, and sensitivity limits with key references. Here HL denotes the Heisenberg limit, and SQL denotes the standard quantum limit, as given in Sec.~\ref{sec-qfi}. While classical systems are bounded by SQL, quantum systems can beat SQL to achieve HL, enabling better precision in sensing. }
\end{table*}}

{\color{black}{\section{Quantum sensing using operator-based Fisher information}

The QFI establishes the ultimate limit on how precisely an unknown parameter can be estimated within a quantum system. In essence, it defines the best possible sensitivity achievable by any measurement protocol, given the quantum state of the system. However, experimentally determining the QFI is a highly nontrivial task, as it generally requires complete knowledge of the quantum state and its explicit dependence on the parameter of interest. In realistic experiments, the estimation of an unknown parameter $\theta$ is typically performed by  suitably chosen observable operator \( \hat{O} \). The precision of the parameter estimation can be quantified using the operator-based Fisher information (OFI), which effectively captures the signal-to-noise ratio associated with measurements of \( \hat{O} \). This experimentally accessible quantity is defined as $\mathcal{F}_O = \lim_{\delta \theta \to 0} \frac{\left( \frac{d\langle \hat{O} \rangle}{d(\delta \theta)} \right)^2} {\mathrm{Var}(\hat{O})},$ where the variance of the observable is  $\mathrm{Var}(\hat{O}) = \langle \hat{O}^2 \rangle - \langle \hat{O} \rangle^2.$ Here, the numerator represents the sensitivity of the signal, i.e.,  how rapidly the expectation value changes with the parameter while the denominator captures the measurement noise arising from quantum fluctuations. The value of OFI is bounded by QFI with the relation $\mathcal{F}_O(\theta, \hat{O}) \leq \mathcal{F}_Q(\theta),$ \cite{Pezze2019} where \( \mathcal{F}_Q(\theta) \) is the quantum Fisher information. Equality is achieved only when the chosen observable \( \hat{O} \) corresponds to the optimal measurement that saturates the quantum Cramér-Rao bound. Thus, while \( \mathcal{F}_Q \) sets the ultimate theoretical bound on precision, \( \mathcal{F}_O \) serves as a practical figure of merit that can be directly evaluated or approximated in experiments.

In this context,  the charge-density-wave (CDW) operator~\cite{Sahoo2024a, Sahoo2024b} was proposed as an observable that quantifies the occupation imbalance between even and odd lattice sites. It is defined as $\hat{O}_{\text{CDW}} = \sum_i (-1)^i \frac{\hat{c}^\dagger_i \hat{c}_i}{n_f}$, where $n_f$ is the total number of particles in the system. This observable can be directly measured in optical lattice experiments with ultracold atoms. In a quasiperiodic system described by the Hamiltonian $H = -\sum_{j} (c_{j}^{\dagger} c_{j+1} + \text{H.c.}) + h \sum_{j} \cos(2\pi \omega j)\, c_{j}^{\dagger} c_j$, where $\omega$ is the ratio of two consecutive Fibonacci numbers, the CDW order parameter, within the single-particle regime, exhibits a localization--delocalization transition. To estimate the parameter $h$, the OFI corresponding to the CDW operator shows superlinear scaling with system size near the transition point, thereby confirming a quantum advantage achievable through this experimentally accessible observable. Fig~\ref{fig:OFI}(a) presents the modulus of the averaged CDW order parameter as a function of the quasiperiodic potential strength $h$. In the delocalized phase, owing to the extended nature of the quantum states, $|O_{\text{CDW}}|$ gradually decreases with increasing system size $L$, whereas it saturates to a finite value in the localized phase. Fig~\ref{fig:OFI}(b) shows the corresponding OFI, $\mathcal{F}_{\text{CDW}}$, as a function of $h$ for various system sizes. The quantity $\mathcal{F}_{\text{CDW}}$ develops a pronounced peak near the finite-size transition point around $h = 2$. Finally, the finite-size scaling of the OFI near the critical point, shown in Fig.~\ref{fig:OFI}(c), follows the relation $\mathcal{F}_{\text{CDW}} \sim L^{1.5}$, confirming that an experimentally realizable observable can indeed provide a quantum-enhanced precision in the estimation of the parameter $h$. }}

\section{Discussion}
\label{sec-disc}
Quantum-enhanced metrology forms one of the most important avenues by which distinctive quantum features like quantum coherence and entanglement are employed to reach super-SQL error domains.  
Dramatic progress has been achieved on the experimental front. Engineered quantum materials are used to push boundaries, where AMO platforms lead the pack, establishing themselves as ideal grounds for precise quantum simulations. In this review, our focus has primarily been on, though not limited to, recent advances in quantum sensing techniques, strategies, and protocols implemented in AMO platforms, or are closely associated with or are particularly well adapted to the same. In Table 1, we present a summarized version of the quantum metrology platforms and strategies for precision estimation of various uknkown parameters, their scaling behavior, and sensitivity limits with key references. Within ensemble systems, we discuss spin ensembles, light-matter systems, and Bose-Einstein condensates. We discussed the potentiality of paradigmatic quantum many-body models, such as quantum spin chains, Fermi-Hubbard models, Bose-Hubbard models, and non-Hermitian systems as quantum sensing devices. 
We surveyed the utility of quantum phenomena, such as spin squeezing, quantum criticality, quantum phase transition (including second-order, topological, and localization transitions), multicriticality, interferometry, time-crystal, and quantum scar, along with quantum coherence, and entanglement, 
in diverse quantum metrological protocols including thermometry, inertial sensors, gravimetry, magnetometry, and precision clocks. The true challenge on the experimental front lies in the preparation and control of scalable quantum simulators with long coherence time. With steady progress in innovative techniques, a promising future awaits in which quantum sensors are poised to achieve near-optimal precision and broad practical impact. 



\bibliography{ref}
\end{document}